\documentclass[iop]{emulateapj}

\usepackage{natbib}
\usepackage{amsmath, amssymb}
\usepackage{times}
\usepackage{apjfonts, graphicx, graphics}
\usepackage[breaklinks,colorlinks,urlcolor=blue,citecolor=blue]{hyperref}
\usepackage[all]{hypcap}
\usepackage{aas_macros}
\usepackage{subfigure}
\newcommand{\tctff}{{t$_{\rm cool}$/t$_{\rm ff}$}}

\shorttitle{Mass Distribution in Galaxy Cluster Cores}
\shortauthors{Hogan et al.}

\begin{document}

\title{Mass Distribution in Galaxy Cluster Cores}
\author{M.~T. Hogan$^{1,2,\ast}$}
\author{B.~R. McNamara$^{1,2}$}
\author{F. Pulido$^{1}$}
\author{P.~E.~J. Nulsen$^{3,4}$}
\author{H.~R. Russell$^{5}$}
\author{A.~N. Vantyghem$^{1}$}
\author{A.~C. Edge$^{6}$}
\author{R.~A. Main$^{7}$}

\affil{
    $^{1}$Department of Physics and Astronomy, University of Waterloo, Waterloo, ON, N2L 3G1, Canada \\
    $^{2}$Perimeter Institute for Theoretical Physics, Waterloo, ON, N2L 2Y5, Canada \\
    $^{3}$Harvard-Smithsonian Center for Astrophysics, 60 Garden Street, Cambridge, MA 02138, USA \\
    $^{4}$ICRAR, University of Western Australia, 35 Stirling Hwy, Crawley, WA 6009, Australia \\
    $^{5}$Institute of Astronomy, Madingley Road, Cambridge CB3 0HA, UK \\
    $^{6}$Centre for Extragalactic Astronomy, Department of Physics, Durham University, Durham, DH1 3LE, UK \\
    $^{7}$Canadian Institute for Theoretical Astrophysics, University of Toronto, 60 St. George Street, Toronto, ON, M5S 3H8, Canada \\
}

%
%

\begin{abstract}
Many processes within galaxy clusters, such as those believed to govern the onset of thermally unstable cooling and AGN feedback, are dependent upon local dynamical timescales.  However, accurately mapping the mass distribution within individual clusters is challenging, particularly towards cluster centres where the total mass budget has substantial radially-dependent contributions from the stellar (M$_{*}$), gas (M$_{\rm gas}$), and dark matter (M$_{\rm DM}$) components.  In this paper we use a small sample of galaxy clusters with deep Chandra observations and good ancillary tracers of their gravitating mass at both large and small radii to develop a method for determining mass profiles that span a wide radial range and extend down into the central galaxy.  We also consider potential observational pitfalls in understanding cooling in hot cluster atmospheres, and find tentative evidence for a relationship between the radial extent of cooling X-ray gas and nebular H$\alpha$ emission in cool core clusters.  Amongst this small sample we find no support for the existence of a central `entropy floor', with the entropy profiles following a power-law profile down to our resolution limit.  
\end{abstract}


\altaffiltext{*}{
    \href{mailto:m4hogan@uwaterloo.ca}{m4hogan@uwaterloo.ca}
}

\section{Introduction}

Galaxy clusters constitute the peaks of the large-scale structure of the Universe, and as such are invaluable cosmological probes \cite[for a review see][]{Allen11}.  Furthermore, individual clusters are sizeable enough to encompass a representative Universal volume, making them ideal natural laboratories to study physical processes ranging from galaxy evolution through to gas dynamics, feedback from active galactic nuclei (AGN), and beyond.  Knowledge of the total mass and its distribution within each cluster is vital, since the cluster mass function is a critical test of cosmological models whilst many of the internal workings of clusters are tuned to the local acceleration.  

Recently the Hitomi satellite \cite[][]{Takahashi10,Kitayama14} observed the central regions ($\lesssim$60~kpc) of the Perseus cluster \cite[][]{Fabian16}, finding remarkably low levels of turbulence in its intra-cluster medium (ICM).  These data showed the turbulent pressure contribution to be only about 4\% of the thermal support.  The core of Perseus is a highly active feedback system harbouring cavities, shocks, and sound waves \cite[][]{Fabian00,Fabian03,Sanders07}.  It had been suggested that such intense nuclear activity may raise ICM turbulence \cite[][]{Churazov01}, potentially biasing X-ray cluster mass estimates low \cite[e.g.][]{Nagai07}.  The Hitomi result therefore suggests that techniques assuming hydrostatic equilibrium are reasonable for measuring masses on relatively small scales even in active environments. This is supported by earlier work showing that turbulent motions are likely suppressed at the centres of giant ellipticals \cite[][]{Werner09}. The caveat that Hitomi was only able to observe a single cluster before its unfortunately premature demise \cite[see][]{Witze16} should of course be considered \cite[][]{Blanton16}.  Similarly, other sources of non-thermal pressure such as cosmic rays are known to be present in clusters \cite[e.g.][and references therein]{Boehringer88,Kravtsov12,Nelson14}.  Nevertheless, deep grating observations suggest that modest non-thermal support ($<$20\%) appears common in relaxed clusters \cite[][]{Sanders11}.  Low turbulence ($\lesssim$25\%) is also seen in at least some isolated ellipticals \cite[e.g.][]{Werner09,Churazov10}, although the turbulent contribution could be higher in more disturbed galaxies \cite[][]{dePlaa12}.

Cluster mass profiles on various scales are calculated using a variety of techniques; for example using X-ray data \cite[e.g.][]{Vikhlinin06,Allen08,Main15}, weak-lensing \cite[e.g.][]{Kubo09,Hoekstra13,Hoekstra15}, the Sunyaev Zel`dovich effect \cite[e.g.][]{Plagge10,Bleem15,Planck15}, galaxy motions \cite[e.g.][]{Wojtak10}, and stellar velocity dispersions \cite[][]{Fisher95,Lauer14}.  However, each technique has its limitations \cite[see e.g. discussions in][]{Mandelbaum15,Biffi16} and we note in particular that hydrostatic mass estimates at $\lesssim$10~kpc are possible for only the most local clusters \cite[e.g. M87, see][]{Romanowsky01,Russell15}.  In this paper we use a small set of clusters whose mass profiles have prior comparisons available at both small and large cluster-centric radii to develop a technique to determine cluster mass profiles across wide radial ranges.  The techniques described will be applied to a larger sample of clusters in upcoming papers to investigate the mass dependence of AGN feedback and thermal instability in clusters more generally. 

A major facet in the quest to fully understand galaxy clusters is determining the processes that maintain the long term state of settled systems. Absent a heating mechanism, the hot ICM of an isolated cluster should cool at a rate of $\sim$100--1000s M$_{\odot}$~yr$^{-1}$ \citep[i.e. the ``classical cooling flow'', see][]{Fabian94}.  However, a deficit of cooling gas at intermediate temperatures \citep[][]{Peterson03,Sanders11} coupled to much lower than predicted levels of molecular gas \cite[e.g.][]{Edge01, Salome03} and star-formation \cite[][]{O'Dea08, Rafferty08} in cluster centres belies this classical model.  The implication is that some mechanism is retarding the predicted cooling.  In the standard picture, radio-mechanical feedback from AGN jets hosted by the central brightest cluster galaxy (BCG) inflate cavities in the X-ray atmosphere, which rise buoyantly and subsequently drive turbulence, weak shocks, and sound-waves in the ICM.  Dissipation of this turbulence is the most plausible heating mechanism to counteract the expected ICM cooling \cite[see][for reviews]{McNamara07, McNamara12, Fabian12}.  Nevertheless, residual cooling must occur within clusters to explain the observed cold gas, ongoing star formation, and extended envelopes of multi-phase gas seen surrounding BCGs \cite[e.g.][]{Edge01,Salome03,McDonald10,Tremblay15} in these so-called `cool-core' (CC) clusters.  

Central to understanding this feedback cycle is determining the physical triggers that initiate transitions between periods of increased heating and cooling.  It has been shown that multi-phase gas and increased star-formation are preferentially observed in cluster cores when the central entropy index $\kappa_{0}$ drops below 30~keV~cm$^{2}$ \citep[][]{Cavagnolo08,Rafferty08,Sanderson09a}, or equivalently the central cooling time t$_{\rm cool}$ drops below 5$\times$10$^{8}$~yrs.  However, outliers show that additional physics is required beyond these simple binary indicators. An intriguing possibility that has recently gained support both from simulations \citep[e.g.][]{McCourt12,Sharma12b,Li15} and observations \citep[][]{Voit15a,Voit15b}, is that the hot gas becomes thermally unstable when the minimum ratio of the cooling time to the free-fall time (\tctff) falls below a certain value,  reported to be approximately 10.  In these `precipitation models', condensation of hot gas ensues once this cooling threshold has been breached.  Alternatively, motivated by low molecular cloud velocities observed by ALMA \cite[][]{Russell14,McNamara14,Russell16,Vantyghem16}, \citet[][]{McNamara16} proposed a model of `stimulated feedback' in which partially cooled gas from cluster cores is triggered to condense by being lifted in the wake of X-ray cavities to an altitude at which its cooling time to infall time {t$_{\rm cool}$/t$_{\rm I}$} falls below some threshold \cite[see also][]{Voit16}.  Both these models vitally depend on dynamical timescales that require robust cluster mass profiles.  Being able to test these models in order to further our understanding of the feedback cycle within galaxy clusters therefore serves as the major motivation for this work.

The paper is arranged as follows.  In Section \ref{Section:Sample} we describe our sample and data reduction.  In Section \ref{Section:ClusterProperties} we investigate the cooling properties of the gas within these clusters.  Section \ref{Mass_Profiles} uses mass tracers at a variety of wavelengths to develop a method for determining mass profiles that is applicable to less well studied clusters and discusses properties that can affect, or be affected by, these mass distributions. Section \ref{Section:FinalRemarks} makes some final remarks, before conclusions are given in Section \ref{Section:Conclusions}.  In this paper we have assumed a standard $\Lambda$CDM cosmology with: $\Omega_{\rm m}$=0.3, $\Omega_{\rm \Lambda}$=0.7, $H_{\rm 0}$=70 km~s$^{-1}$~Mpc$^{-1}$.

\section{Sample and Data Reduction} \label{Section:Sample}

\subsection{Sample Selection}

\begin{table*}
  \centering
  \begin{tabular}{ccccccccccc}
  \hline\hline
    Cluster &   z    & Scale    &  Observation IDs            & \multicolumn{2}{c}{Total Exposure (ks)} &    $N_{\rm H}$         &  \multicolumn{2}{c}{Cluster Centre}  \\
            &        & (kpc/'') &                             & Raw               &   Cleaned           & (10$^{22}$cm$^{-2}$)    &    RA (J2000)  &  DEC (J2000)  \\
  \hline                                                                                                                                             
    A2029   & 0.0773 & 1.464    & 891, 4977, 6101             &  107.63            &  103.31            &    0.033              & 15:10:56.077   & +05:44:41.05  \\
    A2199   & 0.0302 & 0.605    & 10748, 10803, 10804, 10805  & 119.87             & 119.61             &    0.039              & 16:28:38.245   & +39:33:04.21  \\
    A496    & 0.0329 & 0.656    & 931, 3361$^{*}$, 4976        &  104.00            &  62.75             &    0.040              & 04:33:37.932   & -13:15:40.59  \\
    A85     & 0.0551 & 1.071    & 904, 15173, 15174, 16263, 16264 & 195.24         & 193.64             &    0.039              & 00:41:50.476   & -09:18:11.82  \\
    Hydra~A & 0.0550 & 1.069    & 4969, 4970                  &  195.74            &   163.79           &    0.043              & 09:18:05.681   & -12:05:43.51  \\
  \hline
  \end{tabular}
  \caption{Chandra data used in our analysis. Given scales are the angular scale on the sky at the given redshifts using standard cosmology.  $^{*}$ This observation was completely affected by a large flare and removed from analysis.}
  \label{Observations_Table}
\end{table*}

One motivation for our work is to develop a technique to calculate cluster mass profiles in order to study the relevance of the \tctff~ threshold to the onset and magnitude of ICM cooling from the hot phase in cool core clusters.  The techniques described herein are to be applied to a larger sample to study this threshold using more powerful statistical tests in upcoming papers. For now, the aim is to develop the techniques on a smaller selection of well-studied objects where comparison checks to their mass profiles, particularly at small radii, are possible.  

We select the four clusters (A2029, A2199, A496, A85) for which \citet[][]{Fisher95} measured velocity dispersion profiles across a range of radii, and that have $>$100~ks archival Chandra data.  These velocity dispersions allow a comparison to the inner mass profiles.  All four of these clusters were included in the sample of \citet[][]{Main15}, who derived total mass estimates for them.  A2029 has additional X-ray derived total mass estimates from \citet[][]{Allen08} and \citet[][]{Vikhlinin06}, whereas A2029, A2199 and A85 have weak-lensing mass estimates available \citep[][]{Cypriano04,Kubo09}.  To these four we add the Hydra~A cluster\footnote{We hereforth refer to this cluster as `Hydra~A', although note that this name is also used in the literature to refer only to the central radio source and/or BCG.  This cluster is also often erroneously referred to as A780 in the literature -- that cluster is actually a background source with no physical connection to Hydra~A.  Note that `Hydra~A' is distinct from the `Hydra Cluster' (A1060).}, which has dynamical mass constraints at small radii from \citet[][]{Hamer14} in addition to X-ray and weak-lensing mass estimates at higher radii \cite[][]{Okabe12}. 

Each of these clusters has a central cooling time less than 1~$\times$~10$^{9}$~years \cite[][]{Cavagnolo09}.  All have been observed for the presence of H$\alpha$ emission, with detections in all but A2029 \cite[][]{Crawford99, McDonald10}, as well as having had measurements taken of their molecular gas content \cite[][]{Edge01,Salome03}.

\subsection{Data Reduction}

Archival Chandra imaging data for our sample clusters were downloaded from the online repository.  Data were reduced using \textsc{ciao} version 4.7, with \textsc{caldb} version 4.6.7 \cite[][]{Fruscione06}.  Level-1 events were reprocessed to correct for charge transfer inefficiencies (CTIs) and time-dependent gains.  The events were filtered to remove bad grades, with \textsc{vfaint} filtering used when this mode was on.  Background lightcurves were extracted from level-2 events files on a complementary chip to the one that contained the bulk of the cluster emission.  These lightcurves were filtered using the \textsc{lc\_clean} script provided by M.~Markevitch, to remove time periods affected by background flares.

For sources with multiple observations, the reprocessed events files were reprojected to the coordinates of the observation ID (\textsc{obsid}) with the longest exposure.  Blank-sky backgrounds were extracted for each observation and processed identically to the target files, reprojected to the corresponding position, and normalised to match the 9.5--12.0~keV flux of each observation.  Images were created in the energy range 0.5--7.0~keV for each \textsc{obsid}.  For the purposes of identifying point sources and structure in the ICM, these images were summed and background subtracted for each source.  A map of the point-spread function (PSF) was created from the \textsc{obsid} with the longest exposure for each object.  In conjunction with using this PSF map to correct for PSF degradation away from the pointing centre, the \textsc{wavdetect} \cite[][]{Freeman02} algorithm of \textsc{ciao} was used to detect point-sources in each of these summed images.  These point sources were inspected, and where necessary corrected, in DS9 before being masked out from subsequent analysis.  We similarly masked clear structures within the ICM such as cavities and filaments, since these regions are typically out of equilibrium and their inclusion could bias derived properties.

\subsubsection{Spectral Extraction} \label{Section:SpectralExtraction}

The cooling instabilities we aim to study typically occur at small ($\lesssim$10~kpc) radii and so we desire finely binned spectra in the central cluster regions.  The example clusters in this preliminary investigation were purposefully chosen to have deep Chandra data, hence our choice of annuli from which to extract spectra is effectively limited by resolution rather than number of counts.  

Concentric circular annuli were defined for each cluster, centred at the positions given in Table \ref{Observations_Table} (see Section \ref{CentreSection} for cluster centring considerations).   The central annulus is given a width of just 3 pixels (roughly equivalent to the maximal Chandra angular resolution), where each pixel is 0.492~arcsec across.  Successive annuli are given widths that increase by 1-pixel each until the sixth annulus, beyond which we make the width of each annulus 1.5 times the width of the preceding one until we reach $\sim$1400 pixels, giving a total of 16 annuli per source.  This sampling ensures that we have 3--6 annuli with radii $<$10~kpc, that our central annulus is at the resolution limit of Chandra and contains $>$3000 net counts, and that we can trace the resulting profiles to large radii.  The geometric increase in annular size means that the number of counts in each successive annulus increases, which was found to ensure more successful deprojection (see Section \ref{DeprojectionSection}).    

Spectra were extracted separately from each \textsc{obsid}.  Individual response matrix files (RMFs) and auxillary response files (ARFs) were created for each, using \textsc{mkacisrmf} and \textsc{mkwarf} respectively, and the spectra were grouped to a minimum of 30 counts per channel.  The multiple observations for a given source can be separated substantially in time, hence we chose not to sum the spectra, instead keeping them separate and later loading and fitting simultaneously within \textsc{xspec}.  Exposure maps were created for each observation and used to correct for area lost to point sources, chip gaps, etc.

\section{Cluster Properties} \label{Section:ClusterProperties}

\subsection{Projected Profiles} \label{Section:ProjectedProfiles}

\begin{figure*}    
  \centering
    \subfigure[Temperature]{\includegraphics[width=4cm]{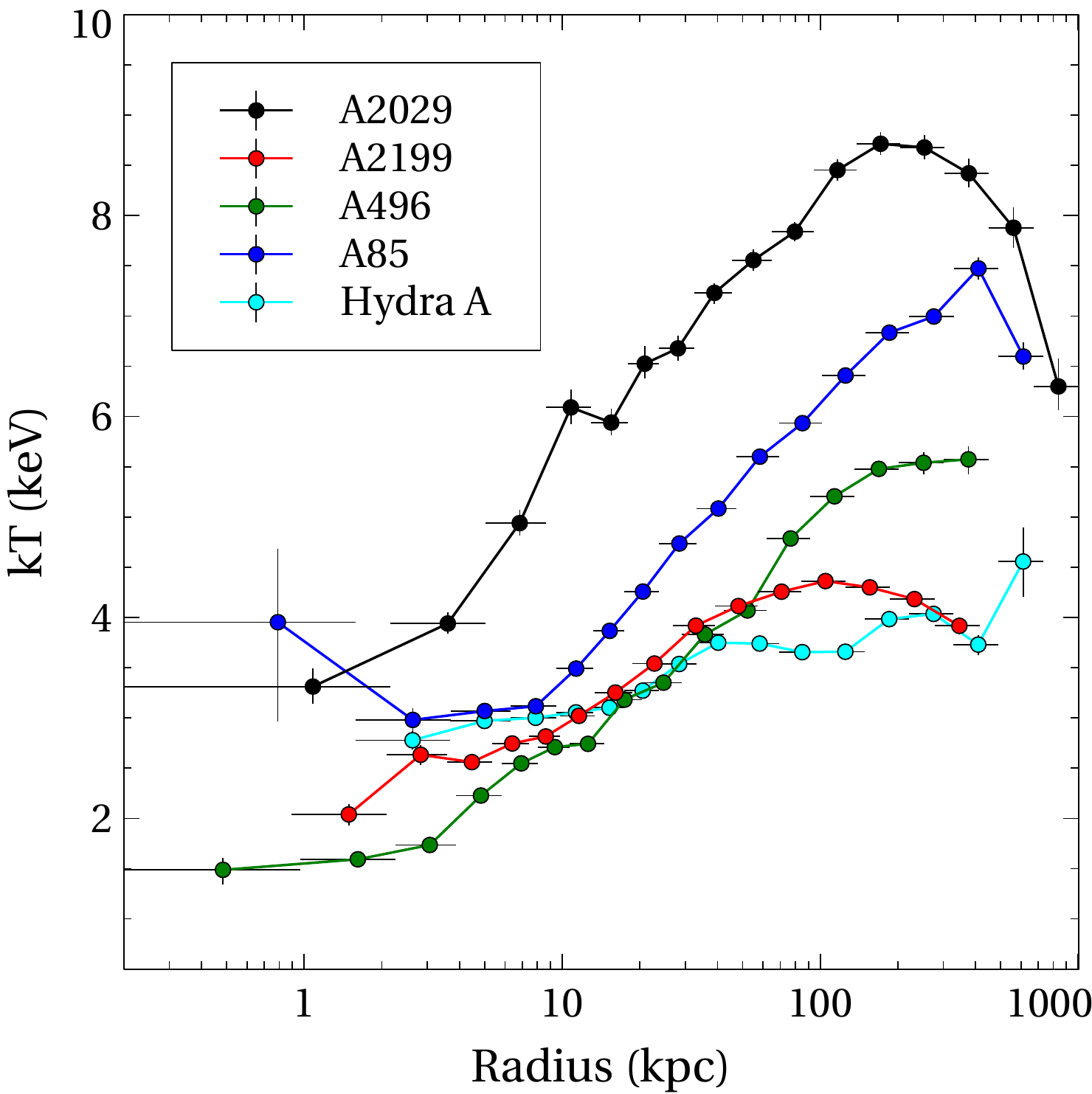}}
    \subfigure[Density]{\includegraphics[width=4cm]{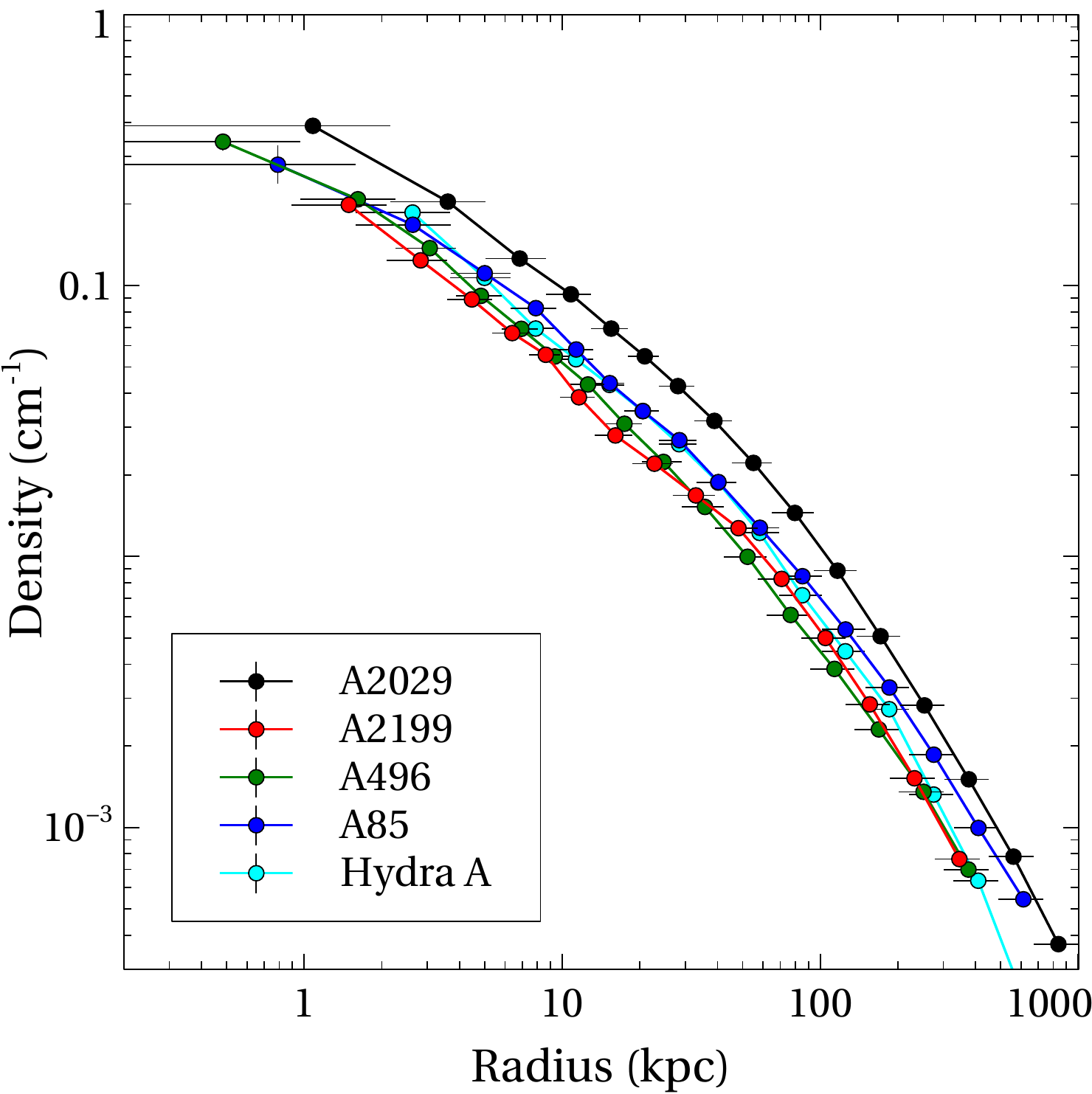}}
    \subfigure[Cooling time]{\includegraphics[width=4cm]{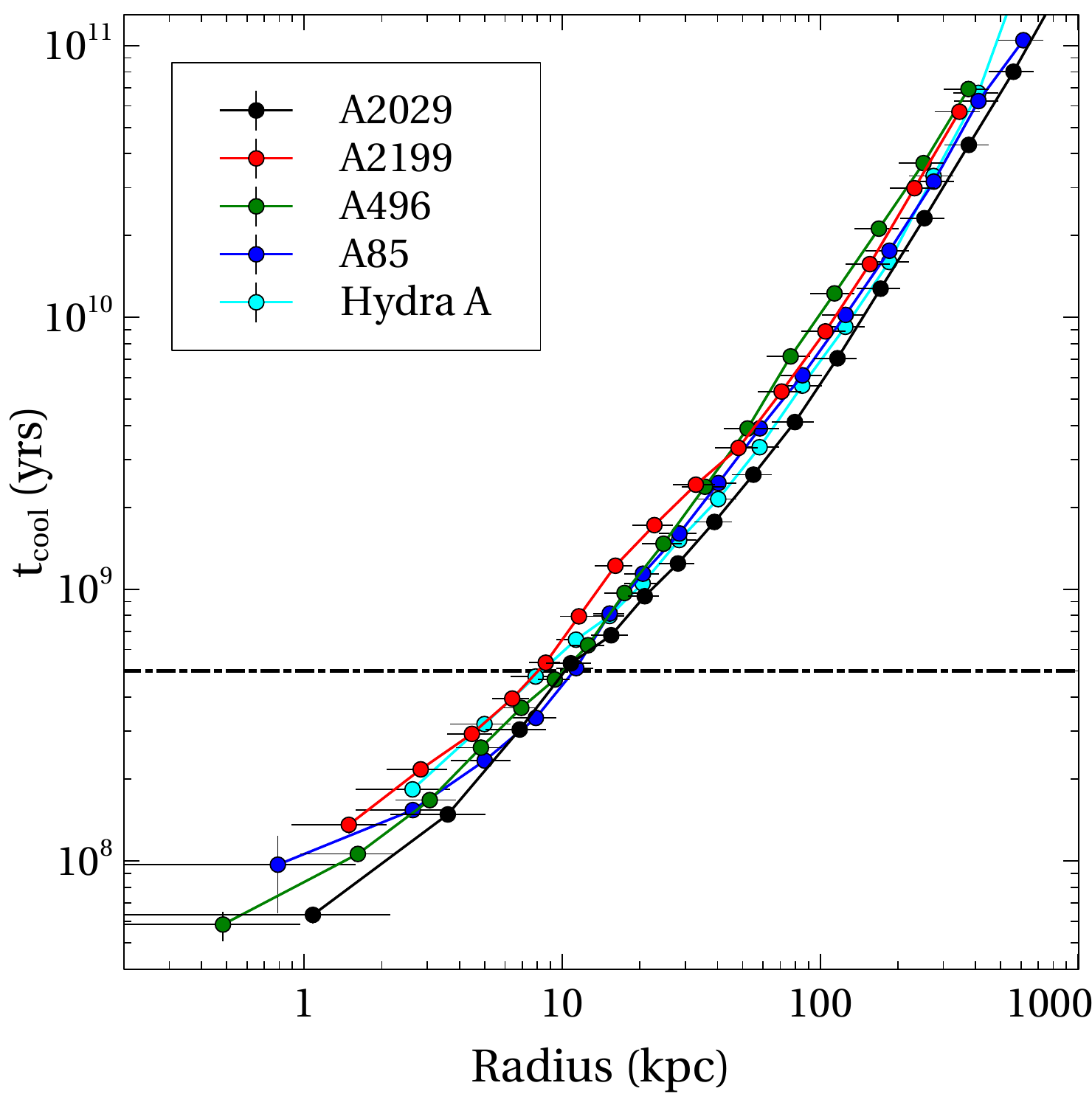}}
    \subfigure[Entropy]{\includegraphics[width=4cm]{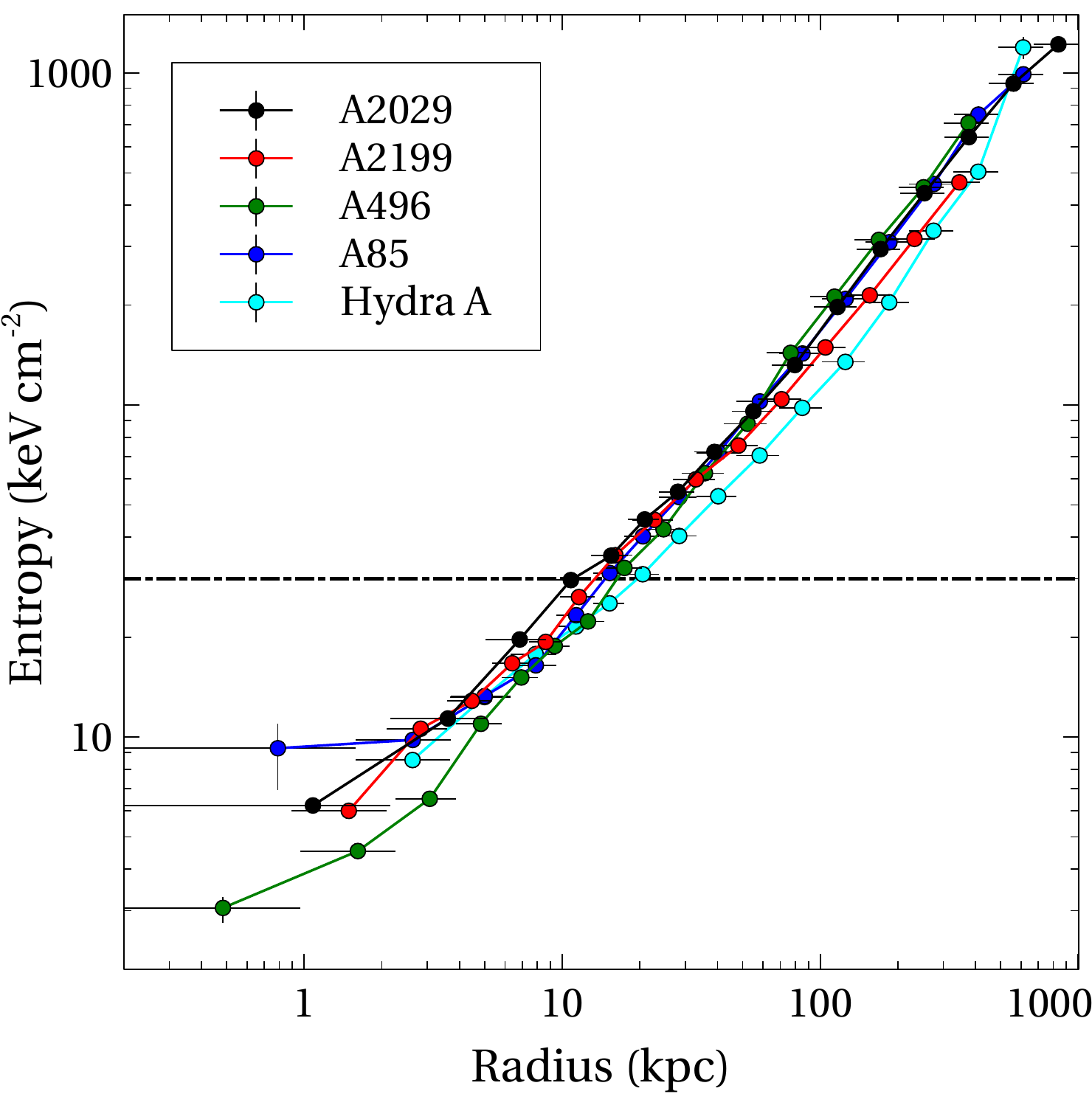}}
  \caption{Projected temperature, density, cooling time (t$_{\rm cool}$), and entropy profiles.  The dotted horizontal lines in panels c and d show the 5~$\times$~10$^{8}$~yr and 30~keV~cm$^{2}$ cooling time and entropy thresholds respectively.  Note that all sources satisfy both the entropy and cooling time thresholds.  Regardless of this and unlike the other four sources shown here, A2029 does not display nebular emission indicative of ongoing cooling.}
 \label{ProjectedProfiles}
\end{figure*}

The extracted spectra for each cluster, alongside their matched response files, were loaded into \textsc{xspec} version 12.8.2 \citep{Arnaud96}.  To derive the projected gas properties of these clusters we initially fitted an absorbed single-temperature (\textsc{phabs}*\textsc{mekal}) model \citep[][]{Mewe85,Mewe86,Balucinska-Church92,Liedahl95}.  Extracted spectra for all \textsc{obsid}s within each annulus were fitted simultaneously.  Our starting values for line of sight galactic absorptions were taken from the LAB Survey \citep[][]{Kalberla05} -- our fitting suggested higher values were required for A2199 and A85, in agreement with the findings of \citet[][]{Main15}, hence we adopted the $N_{\rm H}$ values of these latter authors and subsequently kept these parameters frozen in our analysis (see Table \ref{Observations_Table}).

The outputted temperature and normalisation (N) parameters from our fitted models were used to derive the projected number densities

\begin{equation}
 n_{\rm e}~=~D_{\rm A}(1+z)10^{7}~\sqrt[]{\frac{N~4~\pi~1.2}{V}}
\end{equation}

\noindent where the factor of 1.2 comes from the ionisation ratio n$_{\rm e}$/n$_{\rm p}$.  In turn, these parameters were used to derive pressure ($P~=2~n_{\rm e}~kT$) and entropy ($\kappa_{0}~=~kT~n_{\rm e}^{-2/3}$) profiles.  Cooling times were calculated assuming compressionless thermal cooling, using the relation

\begin{equation}
 t_{\rm cool}~=~\frac{3~P}{2~n_{\rm e}~n_{\rm H}~\Lambda(Z,T)}
\end{equation}

\noindent where P is pressure, n$_{\rm e}$ and n$_{\rm H}$ are electron and hydrogen number densities respectively.  
$\Lambda$(Z,T) is the cooling function for gas at a specific abundance and temperature.  Practically this relation can be simplified to $3~P~V/2~\mathcal{L}_{\rm X}$ where V is the volume $\frac{4}{3}\pi(r_{\rm out}^{3}-r_{\rm in}^{3})$ bounded by the projected edges of each annulus, and $\mathcal{L}_{\rm X}$ its bolometric X-ray luminosity, which we obtain by integrating the unabsorbed thermal model between 0.1 -- 100~keV.  We plot the projected temperature, density, t$_{\rm cool}$, and entropy profiles of our five clusters in Figure \ref{ProjectedProfiles}.  Note that the central annulus of both Hydra~A and A2199 was removed as the emission is dominated by a non-thermal point source associated with their respective AGN.

Of the five clusters studied here only A2029 does not display nebular emission indicative of ongoing cooling, down to an H$_\alpha$ flux limit of $F_{\rm H\alpha}<3\times10^{-16}{\rm erg}~{\rm s}^{-1}~{\rm cm}^{-2}$ \cite[][]{Crawford99,McDonald10}.  In Figure \ref{ProjectedProfiles} we see this cluster has a comparatively higher temperature and density, although note that all five of the clusters satisfy both the central cooling and entropy thresholds in projection.

\begin{figure*}    
  \centering
    \subfigure[Temperature]{\includegraphics[width=4cm]{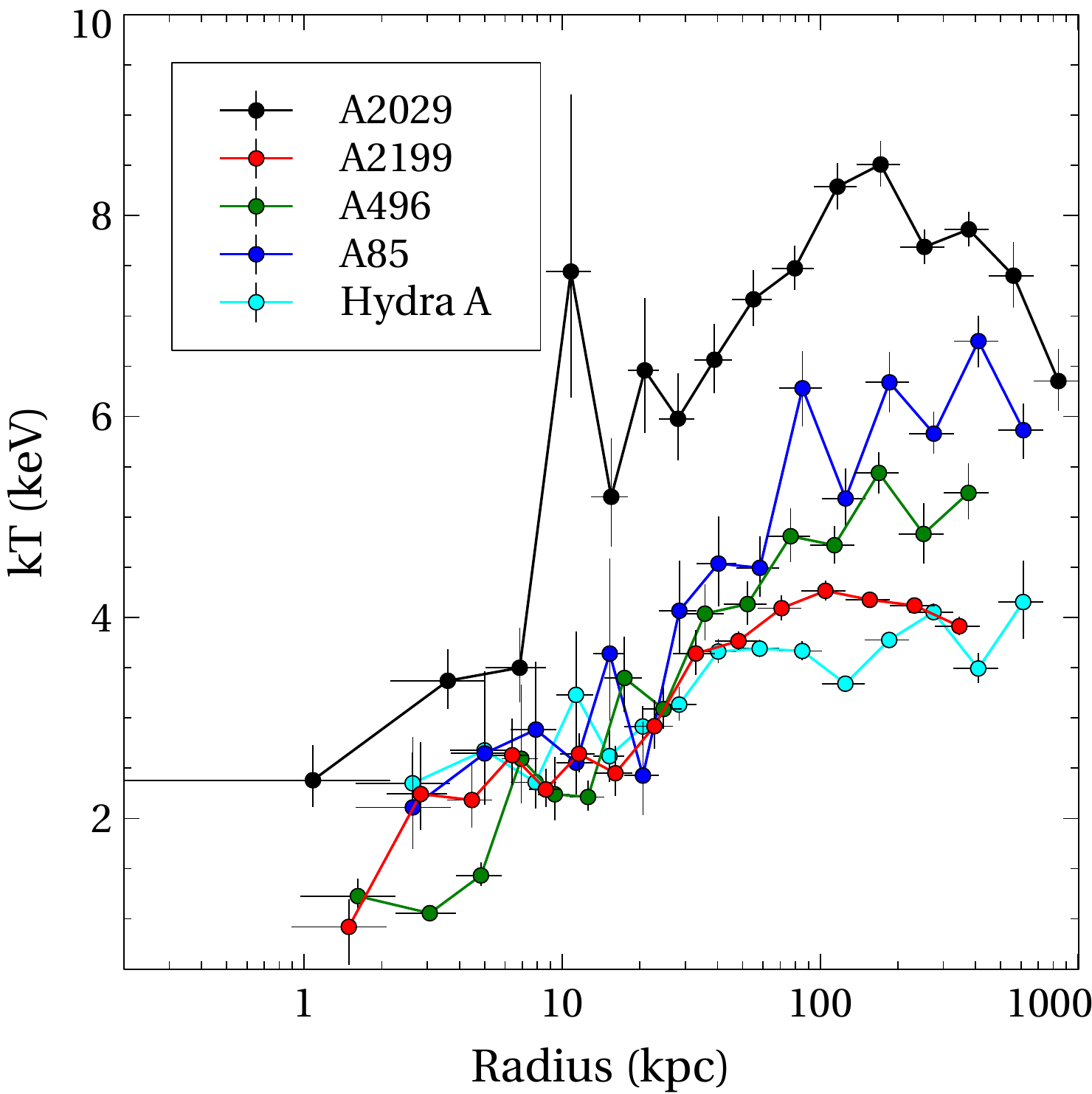}}
    \subfigure[Density]{\includegraphics[width=4cm]{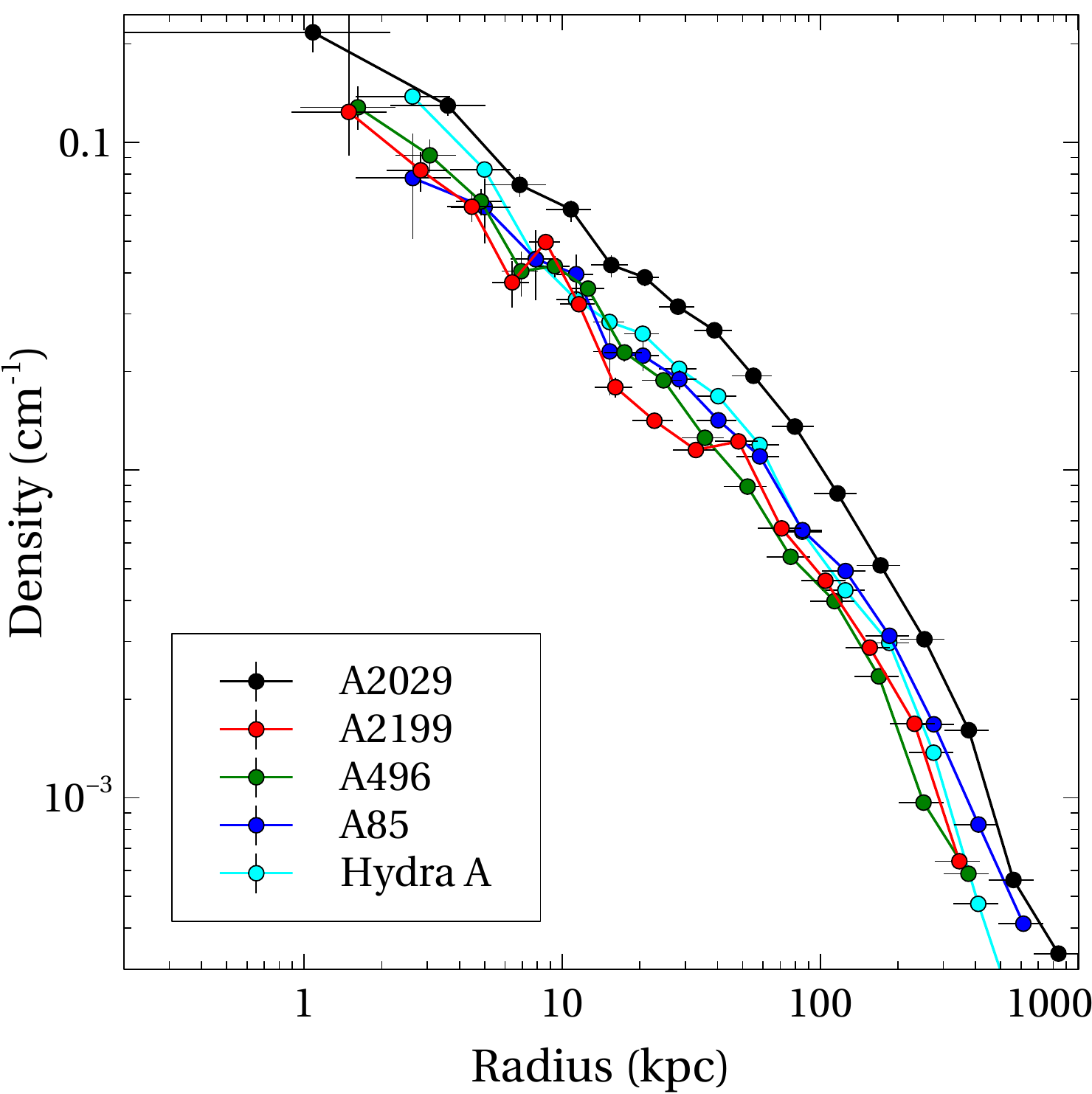}}
    \subfigure[Cooling time]{\includegraphics[width=4cm]{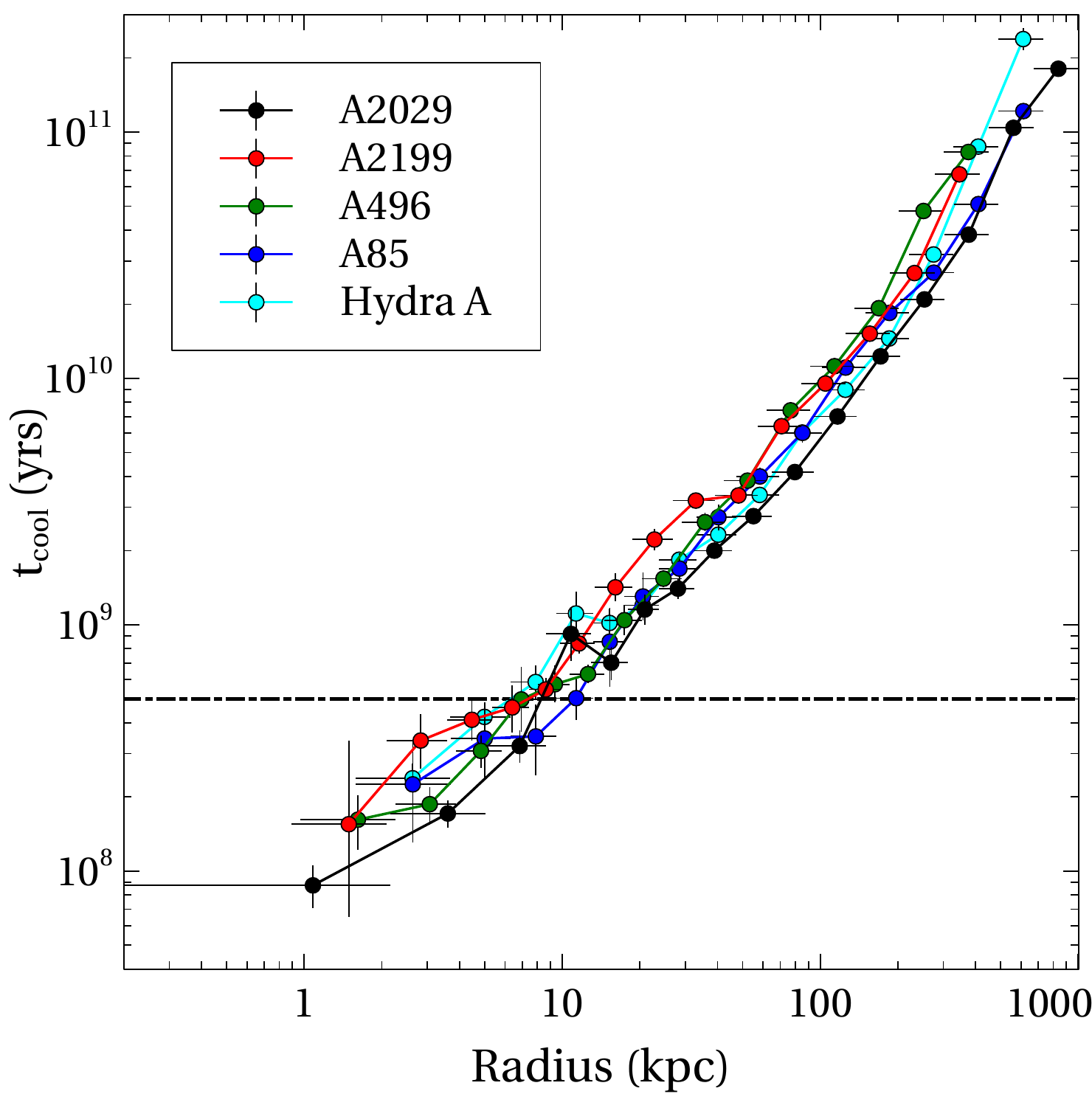}}
    \subfigure[Entropy]{\includegraphics[width=4cm]{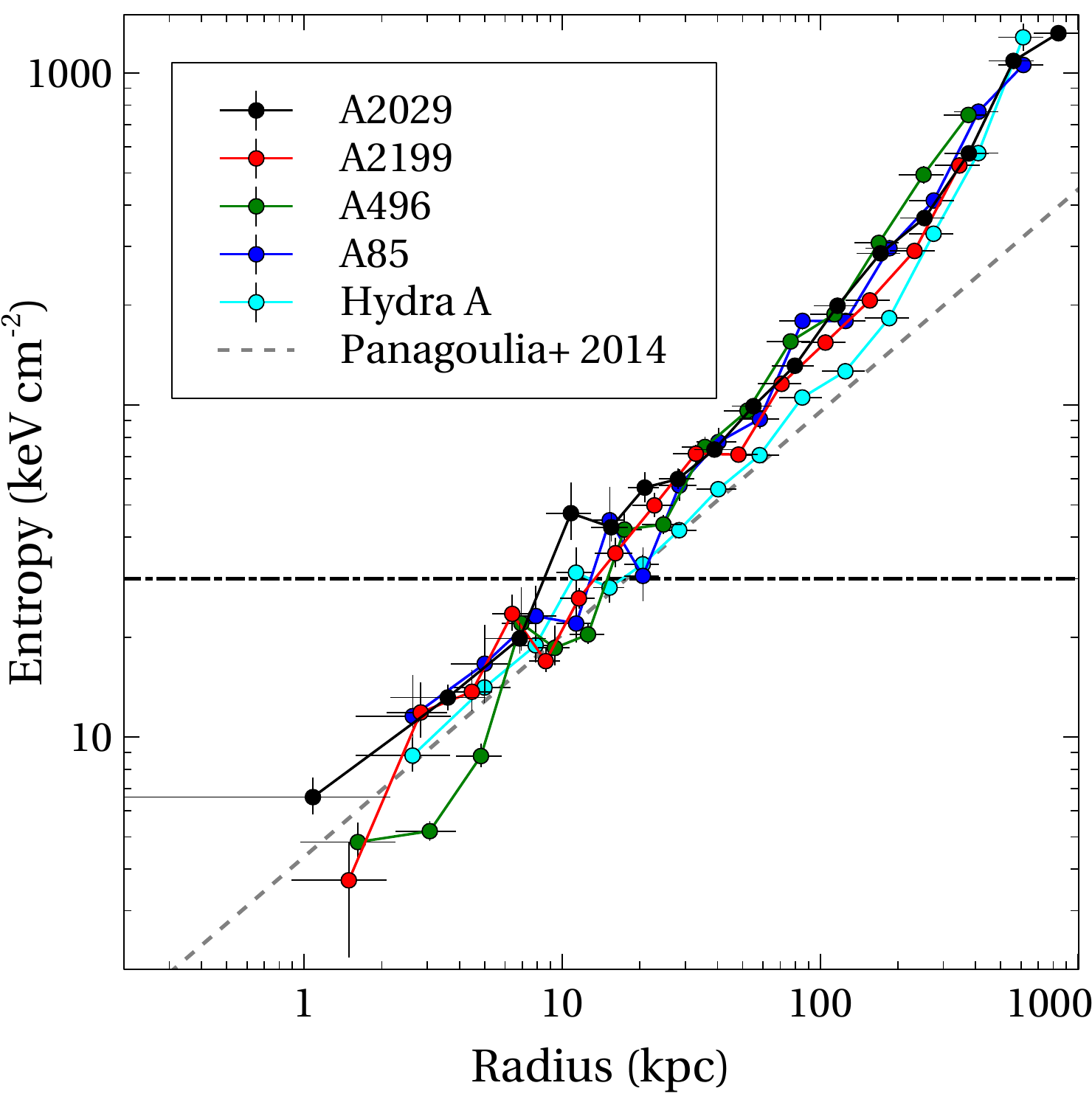}}
  \caption{Fully deprojected temperature, density, cooling time (t$_{\rm cool}$), and entropy profiles.  The dotted horizontal lines in panels c and d show the 5~$\times$~10$^{8}$~yr and 30~keV~cm$^{2}$ cooling time and entropy thresholds respectively.  Note that all sources satisfy the entropy and cooling time thresholds as was the case for the projected profiles (see Figure \ref{ProjectedProfiles}). In panel d we show the overall best fitting power-law model to the inner entropy profiles of a sample of 66 nearby clusters presented by \citet[][]{Panagoulia14}.}  
 \label{DeprojectedProfiles}
\end{figure*}

\subsection{Deprojected Profiles} \label{DeprojectionSection}
Measured central properties of galaxy clusters can be significantly affected by emission from hotter regions at higher altitude contaminating the central spectra in projection.  To derive more accurate profiles therefore requires us to deproject our spectra, which we perform using the model-independent \textsc{dsdeproj} routine (\citealt[][]{Russell08}, also see \citealt[][]{Sanders07,Sanders08}).  Through various trials we found that ensuring both the radius and counts enclosed increased from each inner annulus to the next one out provided the most robust deprojections.

We again fitted an absorbed single-temperature \textsc{phabs}*\textsc{mekal} model to these deprojected spectra and used the fitted quantities to derive the deprojected density, pressure and entropy profiles that are shown in Figure \ref{DeprojectedProfiles}.  Fits failed in the central annulus due to the low number of counts left after deprojection for all sources except A2029, whose central bin contains a physically larger region due to its higher redshift.  The lower central densities derived after deprojection increased the central ($\lesssim$5~kpc) cooling times by $\sim$15--50\%, and the central entropies by $\sim$5--20\%.  All sources still satisfy the central entropy and cooling time thresholds after deprojection.  We note that A2029 has a higher temperature than the other clusters shown here, but that its correspondingly higher density ensures that it meets the cooling criteria.

Our cooling time and entropy profiles continue to fall to small radii. \citet[][]{Panagoulia14} fitted a power-law model to a sample of 66 nearby (z~$<$~0.071) deprojected cluster entropy profiles.  In panel (d) of Figure \ref{DeprojectedProfiles} we show the best fit index and normalisation of their best fitting power-law model.  These authors found no evidence for a flattening of the entropy profiles at small radii (the so-called `entropy floor') as has elsewhere been claimed \cite[e.g.][]{David96,Cavagnolo09,McDonald13}, and suggest that such an effect may be due to resolution.  Our inner entropy profiles agree well with this power-law, and we further support these authors' conclusion that in actuality the entropy profiles may be better approximated by a broken power-law.

\subsection{Two-temperature Fits} \label{Section:TwoTempFits}
\begin{figure*}    
  \centering
    \subfigure[A2199]{\includegraphics[width=4cm]{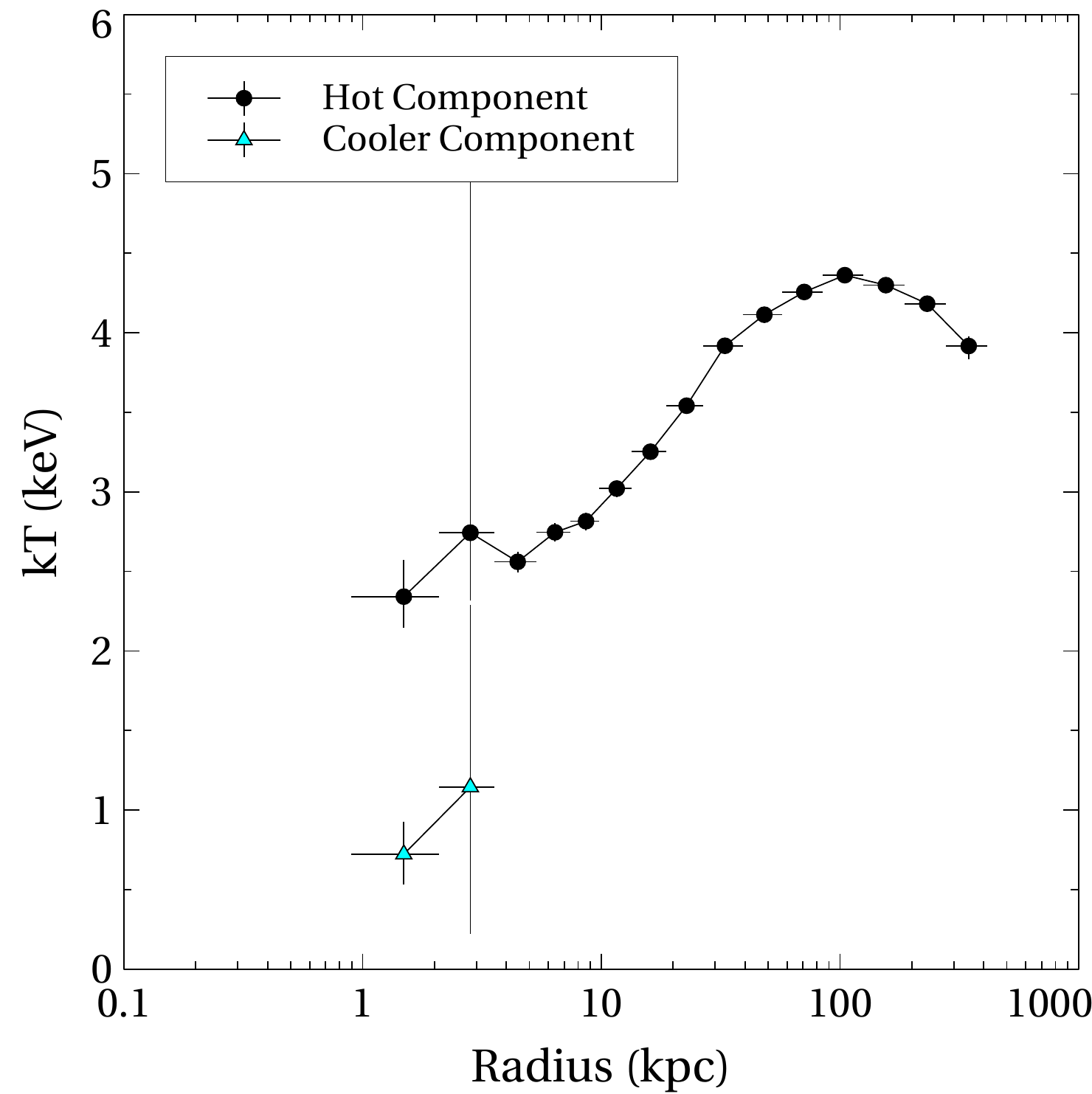}}
    \subfigure[A496]{\includegraphics[width=4cm]{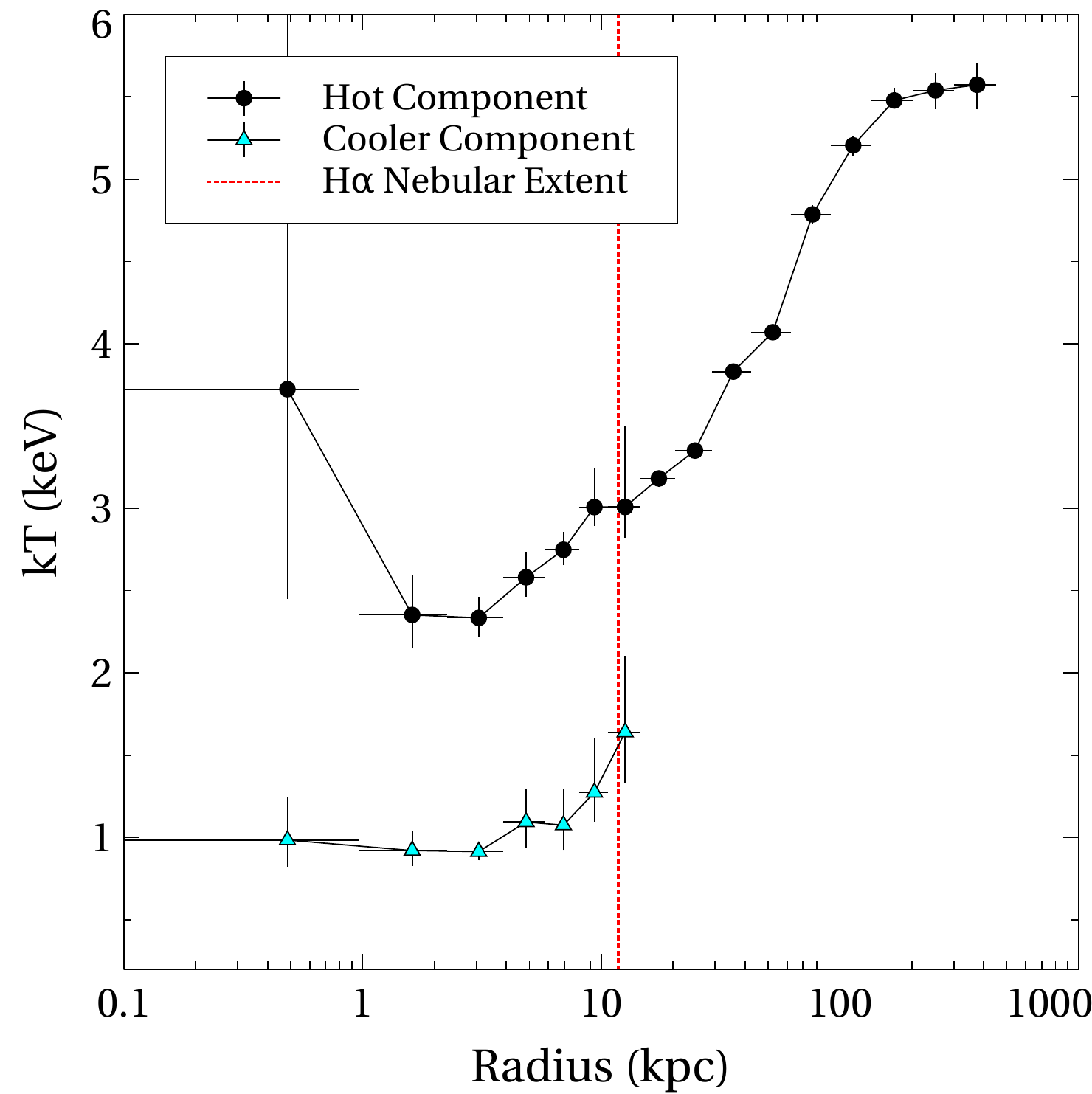}}
    \subfigure[A85]{\includegraphics[width=4cm]{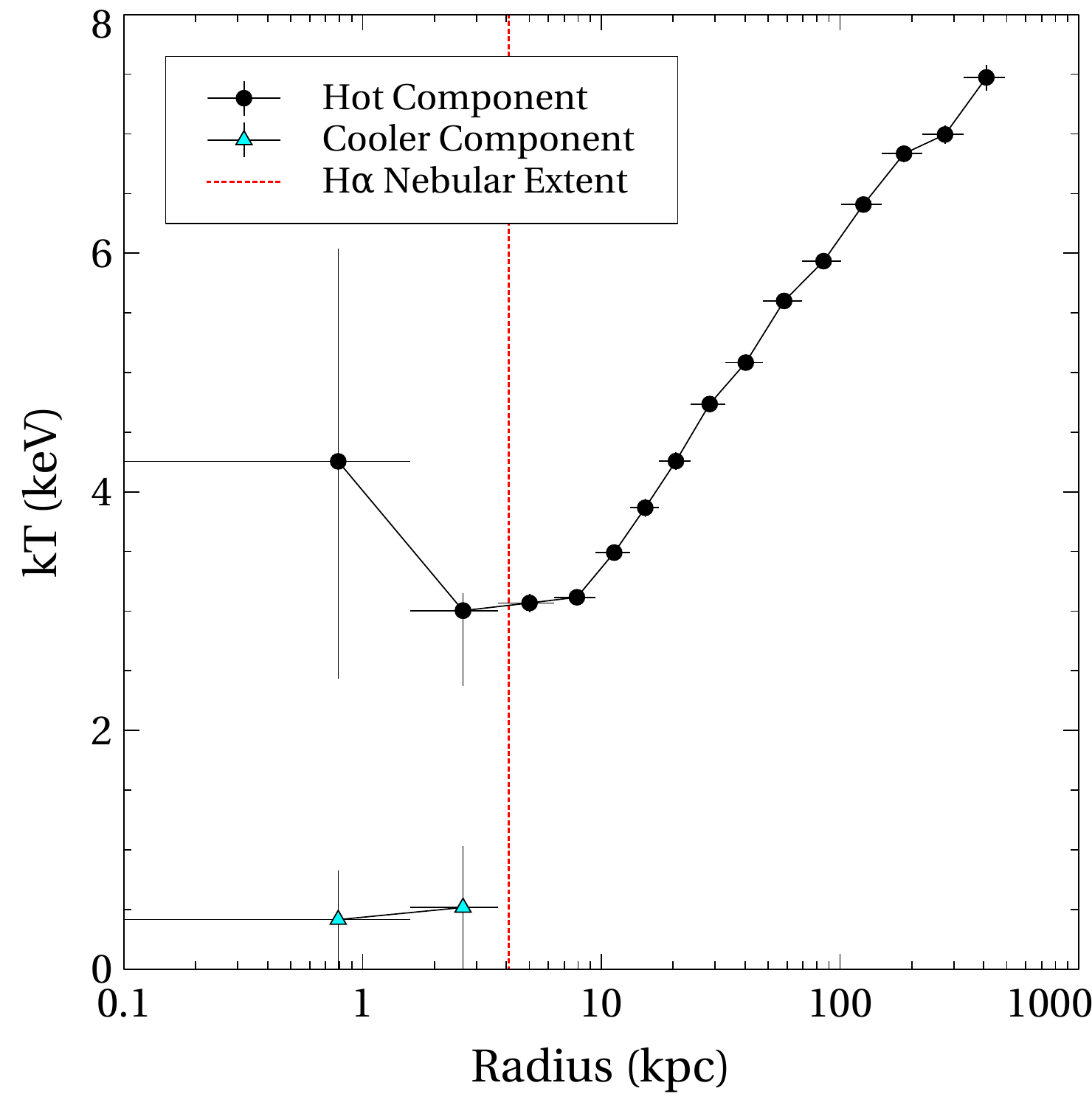}}
    \subfigure[Hydra~A]{\includegraphics[width=4cm]{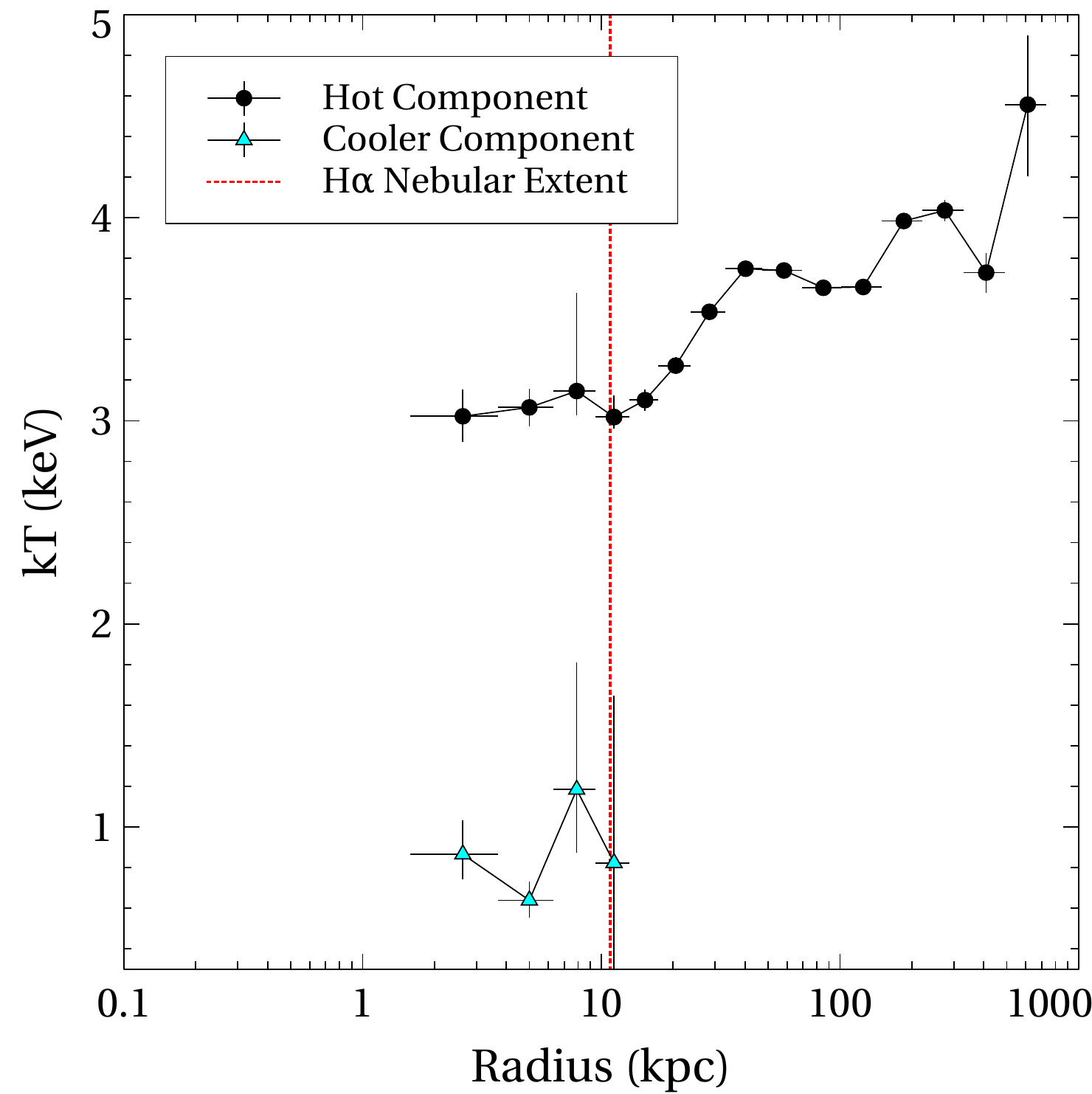}}
  \caption{Projected two-temperature fits showing presence of a cooler temperature component towards the centre of four of our clusters.  Shown in red is the radial extent of the H$\alpha$ nebular emission reported by \citet[][]{McDonald10}.  A2029 showed no requirement for a second temperature component, hence its omission here.  Note that these temperature measurements are likely overestimated -- the resulting counts after deprojection meant that two-temperature models could only be successfully fitted for the projected spectra.  Nevertheless, cooling gas is recovered in all systems with known H$\alpha$.  Furthermore there is tentative evidence for a relation between the H$\alpha$ extent and that of the cooler X-ray component.} 
 \label{2TempProjectedProfiles}
\end{figure*}

The centres of galaxy clusters often contain multi-temperature X-ray gas \cite[e.g.][]{Sanders02,Panagoulia13,Vantyghem14,Russell15}.  To test for this cooling indicator we re-fitted our clusters with a two-temperature model -- \textsc{phabs}*(\textsc{mekal} + \textsc{mekal}).  Applying an F-test, we find that the fits at all radii disfavour the presence of a second temperature component in the ICM of A2029, consistent with the lack of other cooling tracers seen in this system \cite[][]{Edge01,Salome03,Rafferty08,McDonald10}.  For the remaining four clusters we find that two-temperature models are favoured in the inner regions when fitting to the projected profiles, though the extent of the two-temperature region is different for each of the four clusters.

The two-component temperature profiles are shown in Figure \ref{2TempProjectedProfiles}.  Deprojected spectra contained too few counts ($<$8000 in all cases)  to obtain successful two-temperature fits.  This lack of counts is amplified towards the centre (typically $<$500 deprojected counts in central bin), hence the annuli where the strongest two-temperature signal is expected are those that suffer from the lowest number of counts.  Nevertheless, our projected fits show that a two-temperature ICM is present in all four clusters we expect to be cooling, though we caution that the temperatures shown in Figure \ref{2TempProjectedProfiles} are likely over-estimated due to projection.  

\citet[][]{McDonald10} used the Maryland Magellan Tunable Filter to detect and, where relevant, map the spatial distribution of H$\alpha$ nebular emission in a sample of 23 cool core clusters.  Interestingly we find that in the three of our clusters present in \citet[][]{McDonald10} the radial extent of the H$\alpha$ nebula matches the radius to which our fits favour a cooler component within the ICM (see Figure \ref{2TempProjectedProfiles}).  This supports direct cooling from the hot phase as the source of the nebular gas.  Similar spatial correlation between softer X-ray and emission-line nebulae has previously been seen in a small number of objects such as M87 \cite[][]{Sparks04,Werner13} and A426/Perseus \cite[][]{Fabian03}.  However, our small sample precludes any strong conclusions being drawn here and we leave further investigation of this to future study with an extended sample. 

Importantly, we find that the temperature of the hotter component agrees with that recovered from the single-temperature model for radii $>$1~kpc.  Since the hotter component is volume-filling, and we are most interested in the triggering of hydrostatic cooling from this phase, we can calculate cooling profiles of this hotter gas component in a similar fashion to that described in Section \ref{Section:ProjectedProfiles}. For each cluster, the profile derived from the hotter component is in good agreement with that derived from a single temperature fit down to our smallest radii.  Furthermore, although the ratios of the normalisations between the hot and colder gas phases span a relatively wide range they typically fall within $\sim$10-60, showing that the colder component is significantly less abundant ($\lesssim$100$\times$) than the hotter component (see also Section \ref{Extra2TempStuff}).  

Studying NGC~4696 (Centaurus cluster's BCG) \citet[][]{Panagoulia14} found that the presence of a cooler temperature component can suggest flattening in the entropy profile if not fully accounted for.  However, this effect only seems to be present on physical scales $\lesssim$1~kpc in Centaurus and does not appear to flatten our single-temperature entropy profiles (Figure \ref{DeprojectedProfiles}).   Since scales $<$1~kpc are accessible only in the most nearby systems \cite[see][]{Russell15}, single-temperature fits are deemed adequate to trace cluster cooling profiles on the scales of interest to our study (see also Section \ref{Extra2TempStuff}).

\subsection{Resolution Effects} \label{Section:ResolutionEffects}

\begin{figure*}    
  \centering
    \subfigure{\includegraphics[width=17cm]{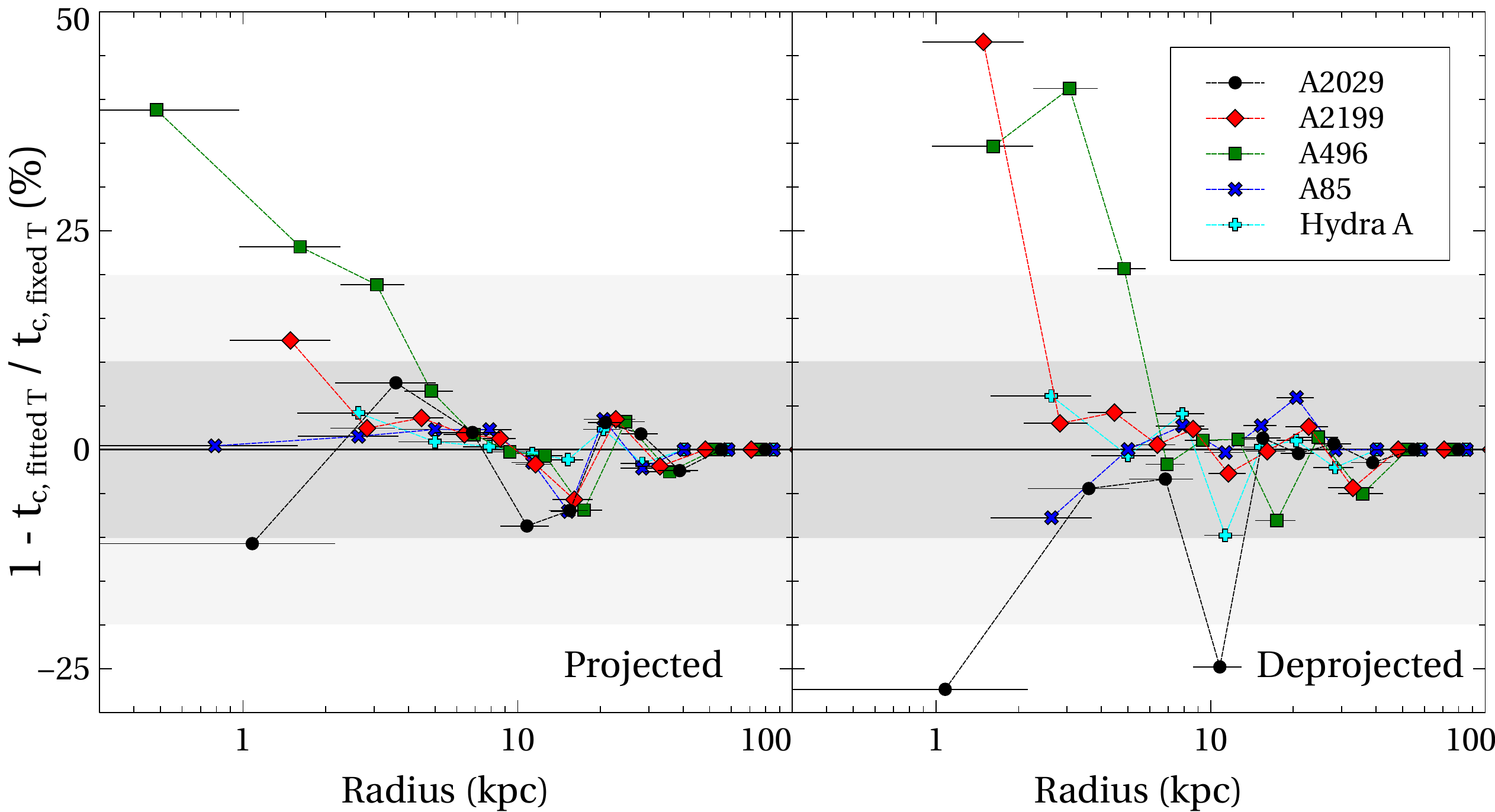}}
  \caption{Percentage error introduced to resulting projected (left) and deprojected (right) cooling time profiles when a single temperature is assumed across the central $\sim$20~kpc, achieved here by first fitting to coarsely binned spectra and then forcing the resulting temperatures on the more finely binned spectra (see text for more details).  Errors are typically $<$10\% in projection, though the effect is exacerbated after deprojection.}
 \label{ResolutionEffects}
\end{figure*}

The clustercentric radius where \tctff~ reaches its minimum, typically $\lesssim$20~kpc \cite[][]{Gaspari12,Voit15a,Voit15b}, is not well resolved in many clusters.  This leads to a variety of effects that can limit the accuracy to which central cluster properties can be measured.  The simplest of these effects is merely redshift related -- the fixed angular resolution of Chandra places an observational lower-limit on the diameter of the innermost annulus, which contains a progressively larger physical volume of cluster core with increasing redshift.  This can bias high the innermost measured temperatures of more distant clusters because the innermost spectrum encompasses typically hotter ICM emission from further out.  Such an effect is unavoidable but must be considered when investigating cooling times amongst a sample of clusters that span a range in redshift.

Another effect that we can attempt to understand here is related to the number of counts required to successfully model the temperature and density of a portion of ICM.  Whilst densities can be reliably determined with only a few hundred counts, a robust single temperature measurement typically requires at least 3000 depending on cluster temperature -- hotter clusters requiring more counts.  Oftentimes this can mean that only a single temperature measurement is possible at $\lesssim$20~kpc, whereas the density profile alone could be measured to much smaller radii.  Determination of the cooling time requires both of these parameters, which leaves an open decision.  Authors can choose to truncate their cooling profiles at the radius required for parallel measurement of temperature and density, can attempt to model and extrapolate the temperature to the inner regions, or can calculate a cooling time profile to smaller radii by using the fitted densities coupled with a constant fixed central temperature \cite[e.g.][]{Cavagnolo09}.  This choice can clearly affect the minimum cooling time.

The extent to which the latter resolution effect ultimately affects cooling times can be investigated by extracting more coarsely binned spectra.  For each of our clusters we extract a single spectrum covering the same region as the innermost few original annuli, such that the central bin extends to an altitude of 18--20~kpc in each case.  We set our second annulus to match the next couple of finely binned regions such that it covers the range $\sim$20--40~kpc, beyond which we revert to using identical regions to those defined in Section \ref{Section:SpectralExtraction}.  

Initially we derive projected and deprojected cooling time profiles for these coarsely binned spectra in the same way as for the more finely binned spectra described in Sections \ref{Section:ProjectedProfiles}--\ref{DeprojectionSection}.  The coarse profiles match our original profiles albeit with higher `central' cooling time due to the higher altitude at which it is measured.  Next, we re-fit the models for the more finely binned spectra but now keep the temperature fixed to that measured across the corresponding coarsely sampled region (i.e. we fix to the corresponding coarse temperature in the inner $\lesssim$40~kpc for each cluster).  The normalisation (hence density) is allowed to vary, thus we effectively simulate what can be done to push cooling profile measurements to small radii in clusters with too few counts to finely trace the central temperature.  The resulting normalisations (alongside the fixed temperatures) are used to calculate `degraded' cooling time profiles.  The percentage difference between our original cooling profiles and these `degraded' profiles are shown in Figure \ref{ResolutionEffects}.

From Figure \ref{ResolutionEffects} we see that cooling time deviations are typically $<$10\% as a result of fixing a central temperature, though the error appears to increase towards smaller radii -- as expected due to greater discrepancy here because the coarsely measured temperature is naturally dominated by emission at higher altitudes within the central bin (here $\sim$20~kpc) due to geometry.  Further, the effect is exacerbated after deprojection.  

An alternative to fixing the inner temperature to a single value is to extrapolate the temperature profile to small radii and then fix to these extrapolated values.  While preferable to a single fixed temperature, we can see by considering panel (a) of both Figures \ref{ProjectedProfiles} $\&$ \ref{DeprojectedProfiles} that the inner temperature profiles of clusters are not necessarily amenable to robust extrapolation.  While density is the dominant driver of cooling time, we caution against over-interpretation of cooling profiles to small radii where projected or coarsely sampled temperatures have been combined with deprojected, more finely sampled densities.

\subsubsection{A further word on two-temperature fitting} \label{Extra2TempStuff}

The same two-temperature models discussed in Section \ref{Section:TwoTempFits} were fitted to the coarsely binned spectra. A second temperature was not favoured by the fits to A2199 nor A85.  Considering Figure \ref{2TempProjectedProfiles} this is perhaps unsurprising -- the cooler temperature component is only present at r~$\lesssim$~3kpc in these systems, hence the $\sim$20~kpc coarse central bin is dominated by single-phase ICM.  However for both A496 and Hydra~A, where Figure \ref{2TempProjectedProfiles} shows the multi-phase region to extend over greater volume, a two-temperature model is recovered in both projection and now also in deprojection -- thus confirming the previous lack of detection in the deprojected spectra as due to the low number of counts in the separate finely binned regions (see Section \ref{Section:TwoTempFits}).

For the projected spectra of Hydra~A (A496) kT$_{\rm h}$=3.04$^{+0.03}_{-0.03}$ (2.97$^{+0.07}_{-0.05}$) keV and kT$_{\rm c}$=0.84$^{+0.15}_{-0.10}$ (1.32$^{+0.10}_{-0.09}$) keV with a ratio of normalisations of $N_{\rm h}/N_{\rm c}$ = 122 (16), where subscripts h and c denote the hotter and cooler X-ray phases respectively.  The corresponding values for the deprojected spectra are kT$_{\rm h}$=2.74$^{+0.07}_{-0.07}$ (2.53$^{+0.05}_{-0.05}$) keV and kT$_{\rm c}$=0.77$^{+0.10}_{-0.10}$ (1.15$^{+0.77}_{-0.85}$) keV with the ratio of normalisations being 84 (22).  

Using $P\propto~n_{\rm e}kT$ and assuming pressure equilibrium between the two X-ray phases, $kT_{\rm h}/kT_{\rm c}=n_{\rm c}/n_{\rm h}$.  The cooler component should therefore be a factor of 3.6 (2.3) denser than the hot component in Hydra~A (A496) respectively.  A consistent ratio is found using either the projected or deprojected spectra in each case.  The \textsc{mekal} normalisation parameter $N\propto~n_{\rm e}^{2}V$, which combined with mass $M\propto~n_{\rm e}V$ gives $M_{\rm h}/M_{\rm c}=kT_{\rm h}N_{\rm h}/kT_{\rm c}N_{\rm c}$.  It follows that the hotter component appears $\sim$450 (40) times more abundant than the cooler component in projection, or $\sim$300 (50) times moreso after deprojection in Hydra~A (A496).  The greater dominance of the hot phase coupled to its flatter temperature profile in Hydra~A compared to A496 likely explains the much higher fractional errors that arise when a single temperature is forced onto the central regions of A496 than when the same is applied to Hydra~A (Figure \ref{ResolutionEffects}).

That the errors introduced by splicing independently fitted temperatures and densities to create a cooling profile appear to be highly dependent on the presence and extent of a second temperature component, as well as the slope of the temperature profile, again suggests great care needs to be employed when attempting to push cooling measurements to small radii.  The measured temperatures for the coarsely sampled hotter component in both Hydra~A and A496 again agrees within 1$\sigma$ with those obtained from a single temperature fit.

\section{Mass Profiles} \label{Mass_Profiles}
\begin{figure*}    
  \centering
    \subfigure{\includegraphics[width=6cm]{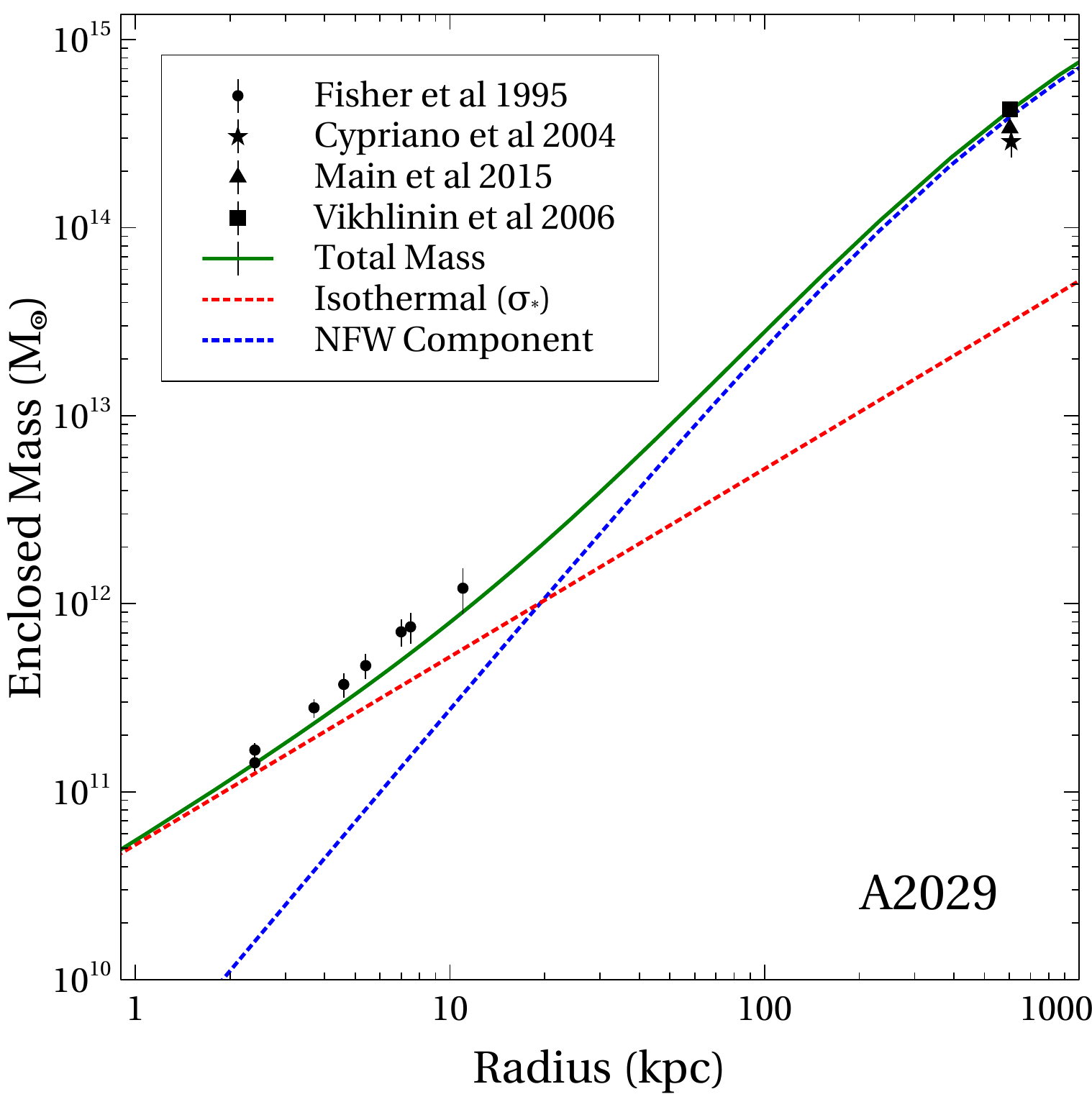}}
    \subfigure{\includegraphics[width=6cm]{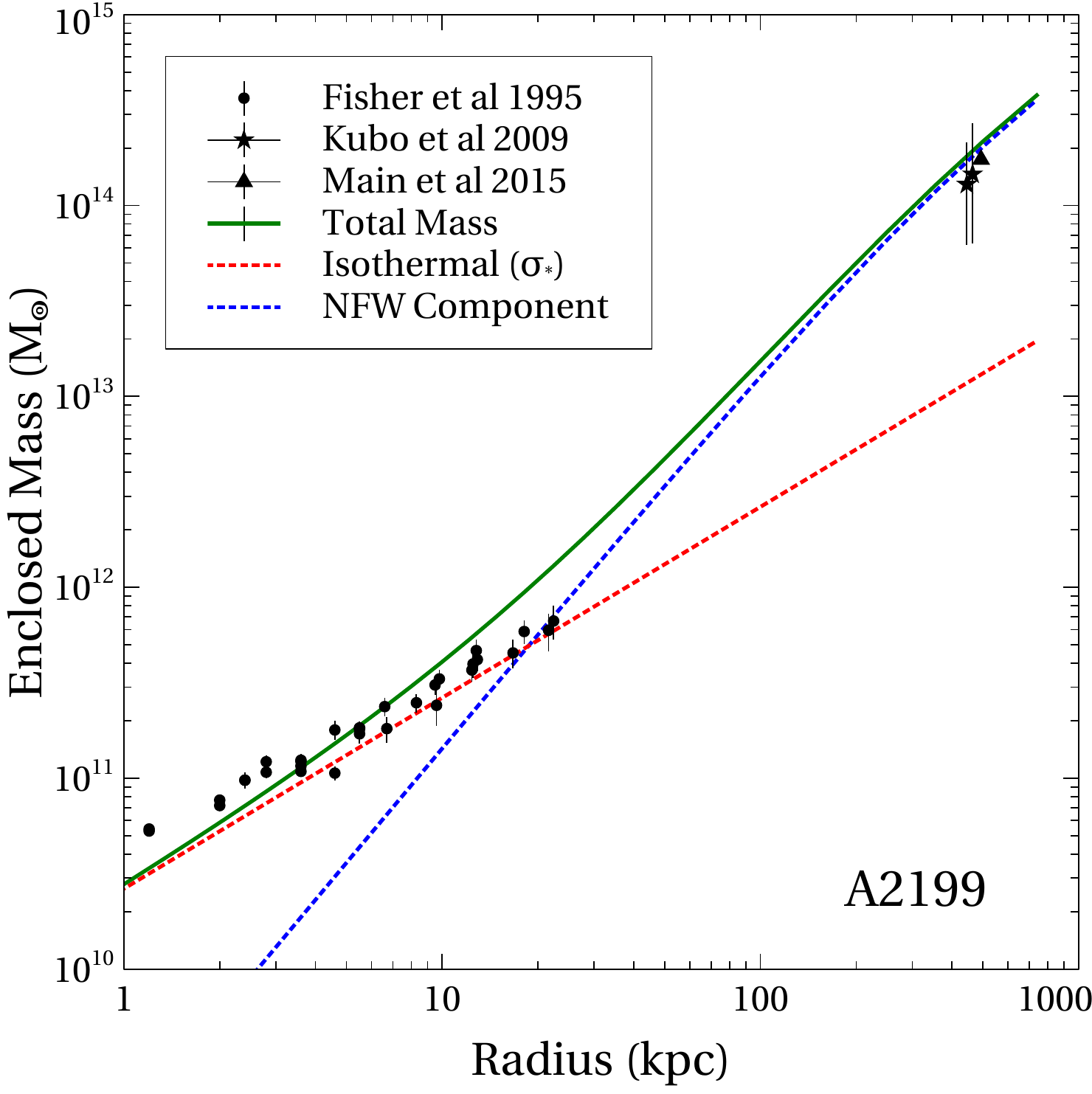}}
    \subfigure{\includegraphics[width=6cm]{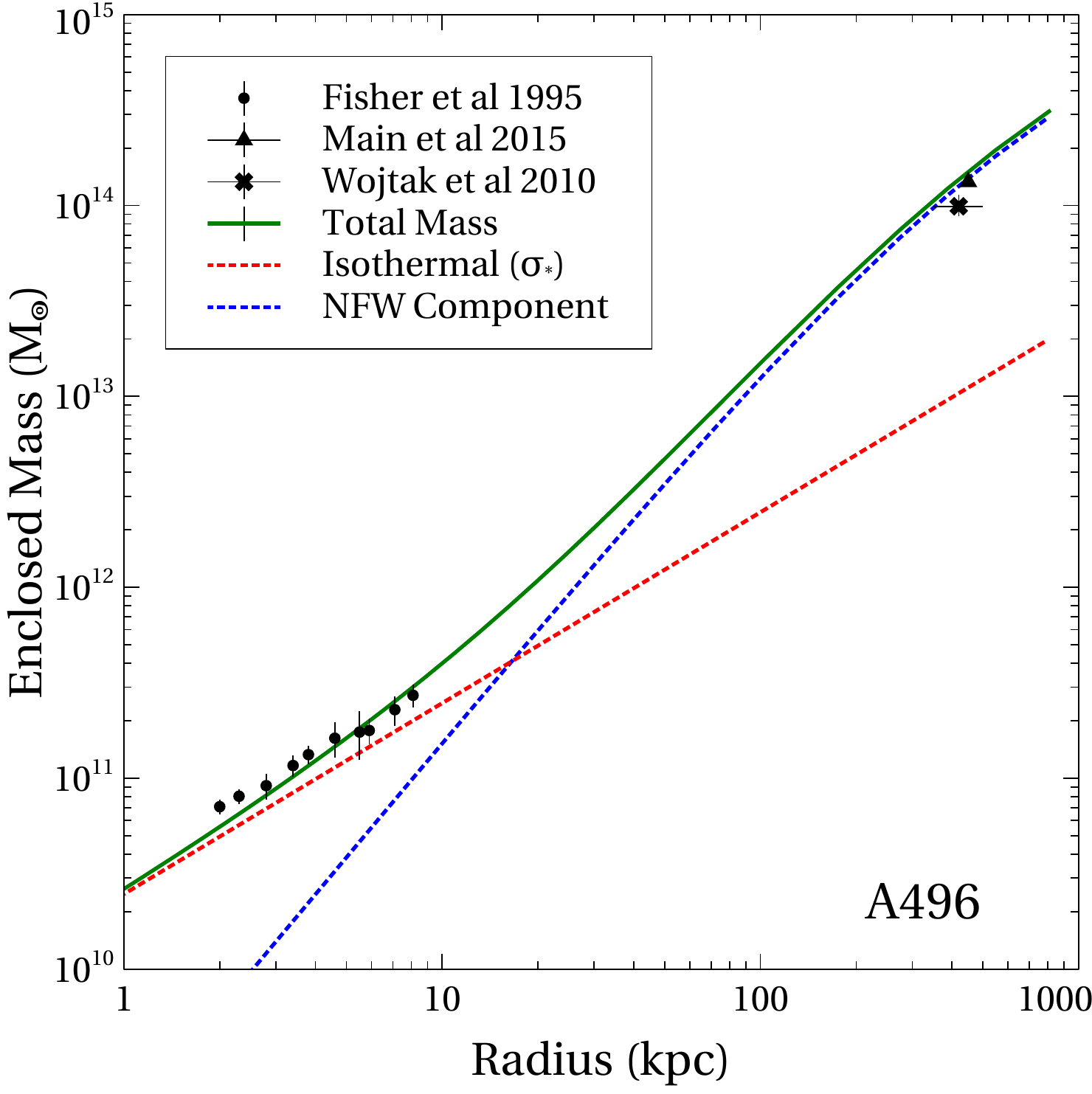}}
    \subfigure{\includegraphics[width=6cm]{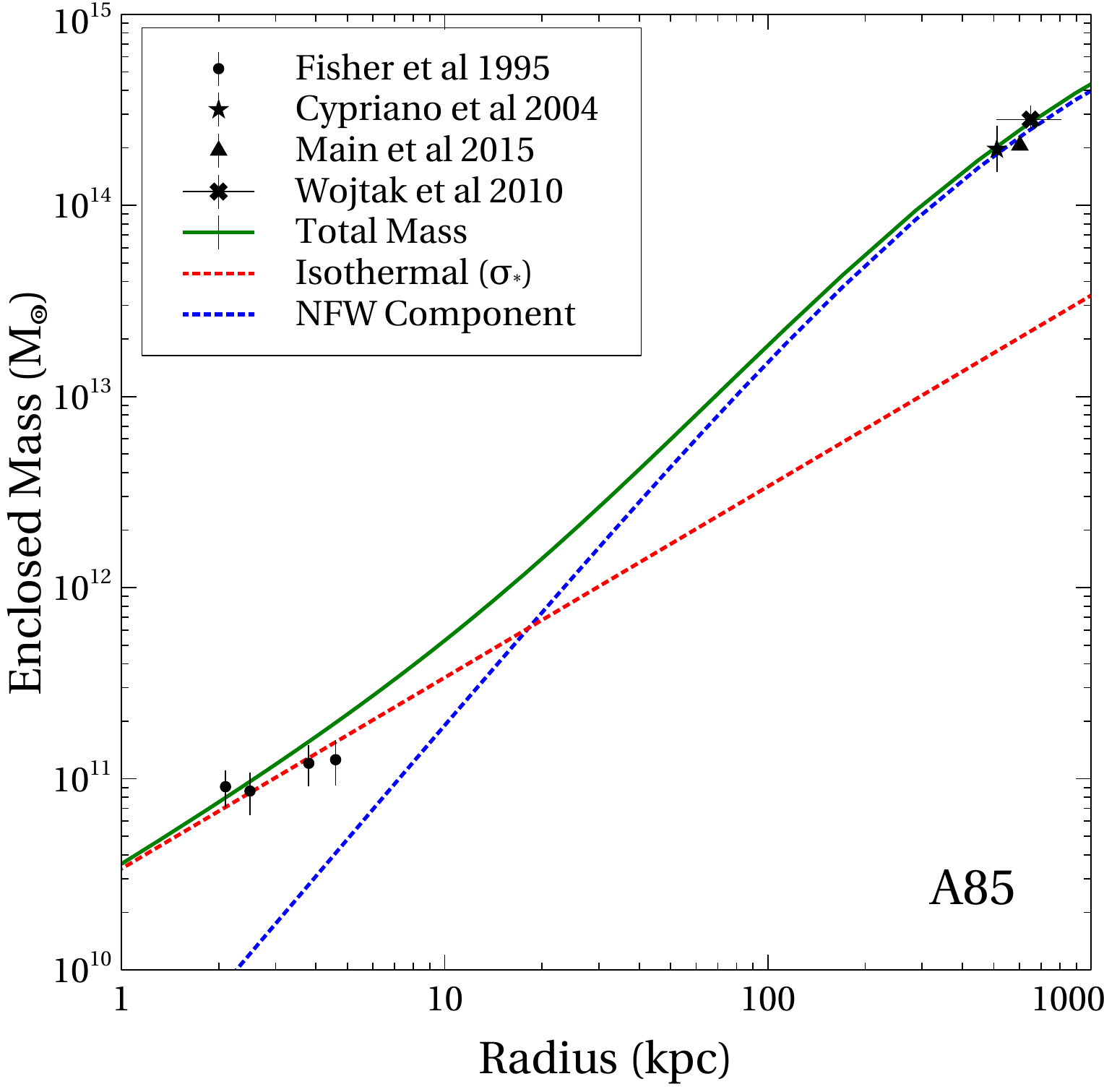}}
  \caption{Mass profiles of the four objects for which \citet[][]{Fisher95} measured velocity dispersion profiles across the BCG.  These profiles consist of an NFW component (blue) to account for the large-scale cluster potential and an isothermal component (red) to account for the stellar components in the inner regions.  Agreement is seen both with the enclosed masses at small radii inferred from the \citet[][]{Fisher95} observations, as well as at large radii where the profiles are compared to the M$_{2500}$ values reported from a variety of sources that employed both X-ray \citep[][]{Vikhlinin06, Allen08, Main15}and weak-lensing \citep[][]{Cypriano04, Kubo09} methods for mass estimation.} 
 \label{MassProfiles}
\end{figure*}

\begin{figure}
	\includegraphics[width=\columnwidth]{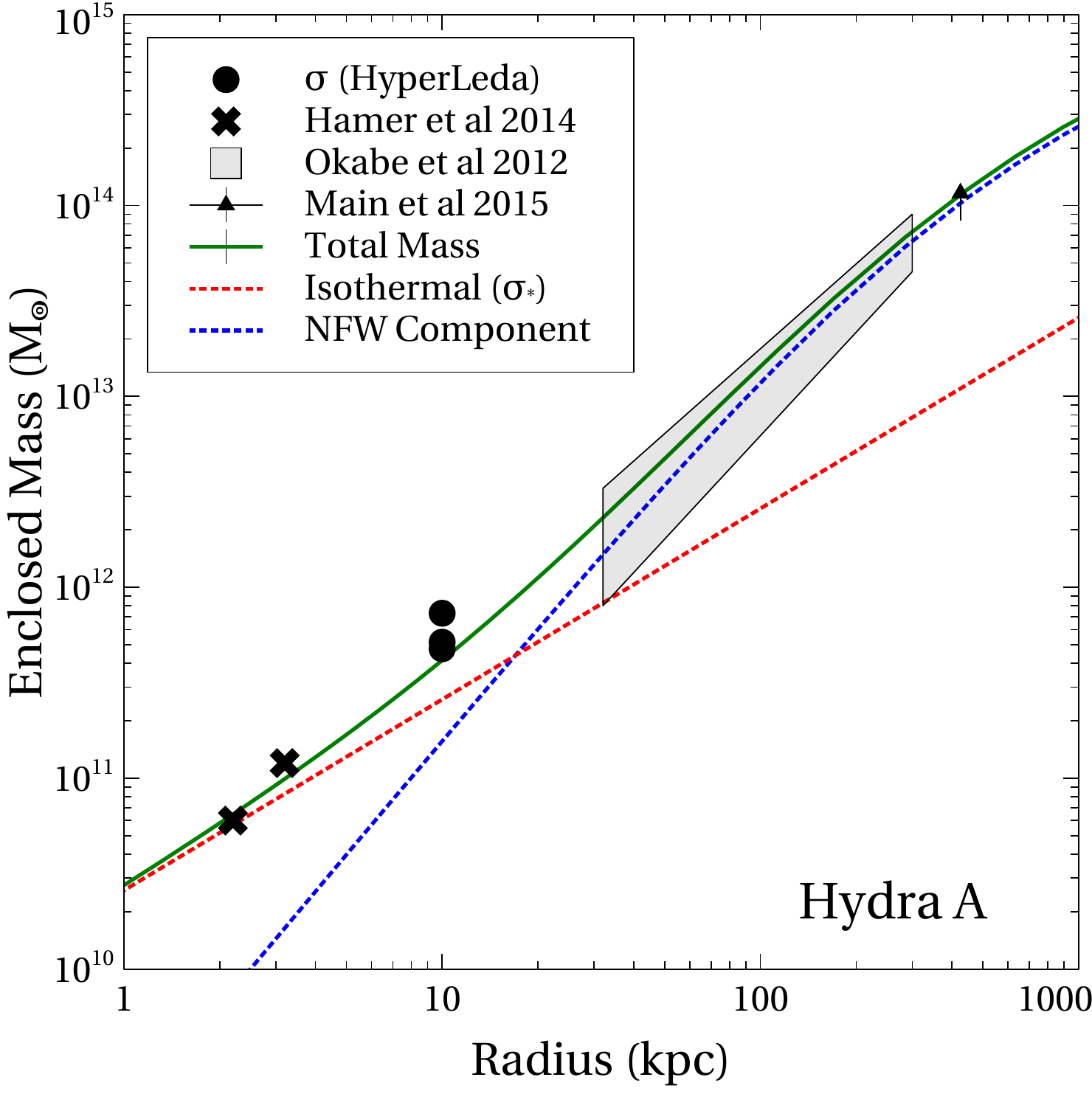}
    \caption{Mass profile for the Hydra~A cluster.  Our combined profile (green) consists of an NFW component (blue) to account for the large-scale cluster potential and an isothermal component (red) to account for the stellar component within the BCG. We achieve good agreement with the total cluster mass (M$_{2500}$) reported by \citet[][]{Main15} from their X-ray analysis, as well as with the weak-lensing derived mass profile of \citet[][]{Okabe12}.  At smaller radii we see agreement with the dynamical masses inferred by \citet[][]{Hamer14} due to the presence of a large molecular gas reservoir.  We additionally show the mass inferred from three velocity dispersion measurements reported in the HyperLeda database.  Note however that these velocity dispersions are reported without their aperture sizes, and hence a typical radius of 10~kpc was taken for plotting purposes.}
    \label{A780_Mass_Profiles}
\end{figure}

In this section we derive mass profiles for our clusters, which are subsequently used to calculate gravitational accelerations and free-fall times.  We briefly outline a couple of approaches that were attempted, discussing their relative merits and shortcomings, before describing our adopted approach.  We present our final derived mass profiles, with comparisons to tracers at other wavelengths and at a range of radii, in Figures \ref{MassProfiles} \& \ref{A780_Mass_Profiles}.  

\subsection{NFW Profiles}
On large scales the NFW profile \cite[][]{Navarro97} is found to be an accurate description of the total gravitating potential of galaxy clusters \cite[e.g.][]{Pointecouteau05,Vikhlinin06,Schmidt07}.  Following \citet[][]{Main15} we initially employed the cluster mass mixing model \textsc{nfwmass} within the \textsc{xspec} package \textsc{clmass} \cite[][]{Nulsen10}, to fit NFW profiles to our chosen clusters.  This model assumes that the X-ray emitting gas is in hydrostatic equilibrium and that the cluster is spherically symmetric.  

Although a single NFW provides a good fit to the global mass profiles of galaxy clusters, at small radii the NFW alone underestimates the masses inferred from stellar velocity dispersions in the central galaxy \cite[e.g.][]{Fisher95, Lauer14}.  An additional component, assumed to be mainly due to the stellar mass of the BCG, therefore needs to be accounted for when fitting to the observational data.  

An understandable approach taken by \citet[][]{Voit15a} to account for the BCGs in their large sample of clusters was to invoke a stellar velocity dispersion floor at 250~km~s$^{-1}$ in each cluster's mass profile.  Whilst this is reasonable on average, and motivates our own investigation, it does mean that all clusters tend towards the same mass profile in the central regions that are most sensitive to the postulated \tctff~ instability.  Effectively this means that there is very little range in the denominator, and so the spread in the \tctff~ ratio amongst the cluster population is almost entirely dictated by their cooling times \cite[see][]{McNamara16}.  It should also be considered that the velocity dispersions of many BCGs can be much higher than 250~km~s$^{-1}$.  For example,  \citet[][]{Lauer14} measured stellar dispersions in all BCGs hosted by Abell clusters at z~$\lesssim$~0.08 and reported only 24.4\% (92/377) as having $\sigma$~$<$~250~km~s$^{-1}$, with some BCGs having much higher dispersions such as 462~$\pm$~39~km~s$^{-1}$ in A193 and 451~$\pm$~10~km~s$^{-1}$ in A2716.  In many cases a floor of 250~km~s$^{-1}$ therefore biases the inner mass and associated \tctff~ low.  Ideally we would like to assign an individually tailored inner mass profile to each cluster.  The nearby \citet[][]{Lauer14} sample aside, velocity dispersions in BCGs are in short supply, hence we desire a method to obtain knowledge of the inner mass distribution that is applicable to all observed clusters.

\subsection{Beyond a Single NFW Potential}
To attempt to directly fit the X-ray data for a second component in the mass profile, the \textsc{nfwmass} models were extended to allow the inclusion of an additional potential alongside the NFW potential

\begin{equation}  \label{Equation:NFWPotential}
\Phi_{\rm NFW}(r) = -4{\pi}G\rho_{0}~r_{\rm s}^{2}~\frac{\ln(1~+~r/r_{\rm s})}{r/r_{\rm s}}
\end{equation}

\noindent where $\rho_{0}$ is the characteristic density and $r_{\rm s}$ the scale radius.  Three options were considered for the second potential, with the resultant models dubbed \textsc{nfwnfwmass}, \textsc{isonfwmass}, and \textsc{hqnfwmass}.  The \textsc{nfwnfwmass} model simply had two concurrent NFW potentials with independent densities and scale radii.  The \textsc{hqnfwmass} and \textsc{isonfwmass} models consisted of an NFW profile combined with a Hernquist and an isothermal potential respectively.  The Hernquist potential is

\begin{equation}
\Phi_{\rm H}(r) = -\frac{GM_{\rm H}}{r_{\rm H}}\frac{1}{1+r/r_{\rm H}}
\end{equation}

\noindent where M$_{\rm H}$ is the total mass assigned to this component and $r_{\rm H}$ is its scale radius.  When a Hernquist profile is used to describe the stellar mass of the BCG this scale radius $r_{\rm H}~=~r_{\rm e}/1.815$, where $r_{\rm e}$ is the effective radius of the BCG \cite[][]{Hernquist90}.  A basic isothermal sphere has potential $\Phi_{\rm iso}(r)$~=~2~$\sigma^{2}~\ln(r)$, which can cause issues at $r=0$.  Instead, a modified cored isothermal potential is used

\begin{equation} \label{IsothermalEquation}
\Phi_{\rm iso,c}(r) = \sigma^{2}\ln(1+(r/r_{\rm I})^{2})
\end{equation}

\noindent where $r_{\rm I}$ is a scale radius and $\sigma$ the velocity dispersion.  The core is a numerical requirement to prevent a singular central potential but can be used to account for the flattened stellar cores seen in many BCGs.

It was found that directly fitting two components to the X-ray data with these mixing models did not behave as hoped.  The non-NFW component in the mass profiles is expected to be mainly attributable to the stellar mass content of the BCG.  The resolution of Chandra coupled to the number of counts required per annulus for our mass modelling means that we only expect one or two annuli to contain an appreciable amount of this second component.  Practically this meant that the mass profile was dominated by the single NFW and that the second component was typically minimised and only unstable solutions could be found.  The \textsc{nfwnfwmass} model in particular was found to be most unstable, with the similar potential seemingly vying for dominance, hence we removed this model from consideration.

\begin{table*}
  \centering
  \begin{tabular}{cccccccccc}
  \hline\hline
    Cluster & $\sigma_{*}$     & $\rho_{\rm 0,ISO}$ &  R$_{\rm s,NFW}$    & $\rho_{\rm 0,NFW}$   & R$_{2500}$ & M$_{2500}$              \\
            &  (km~s$^{-1}$)   &   (keV)       & (arcmin)      & (keV)           &  (kpc)    & ($\times$10$^{14}$~M$_{\odot}$)  \\
  \hline                                                                                                                         
    A2029   & 335.9$\pm$10.0  &   0.694        & 6.79$^{+0.49}_{-0.46}$  &  88.68$^{+4.84}_{-4.10}$ & 686.1  & 4.94$^{+0.17}_{-0.19}$ \\[5pt] 
    A2199   & 238.9$\pm$4.0   &   0.351        & 26.05$^{+2.41}_{-3.07}$ &  72.48$^{+5.43}_{-6.82}$ & 558.1  & 2.54$^{+0.12}_{-0.18}$ \\[5pt] 
    A496    & 228.1$\pm$4.6   &   0.320        & 14.00$^{+2.88}_{-2.09}$ &  45.65$^{+5.93}_{-3.68}$ & 482.5  & 1.65$^{+0.11}_{-0.11}$  \\[5pt] 
    A85     & 270.4$\pm$6.4   &   0.450        & 7.37$^{+0.46}_{-0.21}$  &  49.24$^{+1.64}_{-0.92}$ & 516.7  & 2.07$^{+0.04}_{-0.03} $ \\[5pt] 
    Hydra~A & 236.6$\pm$8.4   &   0.344        & 5.85$^{+0.53}_{-0.49}$  &  32.29$^{+1.52}_{-1.55}$ & 423.6  & 1.14$^{+0.03}_{-0.04}$  \\[5pt] 
  \hline
 \end{tabular}
  \caption{Details of the \textsc{isonfwmass} profile fits.  Columns are: i) Cluster name, ii) equivalent stellar velocity dispersion, iii) isothermal potential = ${\mu}m_{\rm H}\sigma^{2}$ where $m_{\rm H}$ is the mass of the hydrogen atom and the mean atomic weight $\mu$=0.59, iv)  NFW scale radius, v) NFW potential = $4{\pi}G{\rho}_{0}R_{\rm s}^{2}{\mu}M_{\rm H}$ in units of keV, vi) R$_{2500}$, vii) M${2500}$.  The reported $\rho_{\rm 0,ISO}$ values correspond to the $\sigma_{*}$ values and were kept fixed in the fitting to account for the anchored stellar mass component.  See text for more details.}  
  \label{ISONFW_fits_table}
\end{table*}

\subsection{Anchoring the BCG Stellar Component}
Anchoring the second component to an observable that is available for all (or at least the vast majority of) BCGs whose clusters have been or are likely to be observed with Chandra is preferable, as this will allow an approach that can be applied homogeneously to a large sample.  The 2-Micron All Sky Survey \cite[2MASS,][]{Skrutskie06} covers essentially the full sky at J, H, and K-bands.  The K-band (2.17$\mu$m) extended source catalog is well-suited to our needs since this wavelength is a good tracer of stellar light whilst not being overtly affected by AGN and dust emission.

It was highlighted by \citet[][]{Lauer07} that the relatively shallow survey depth of 2MASS means that the size of the extended envelopes known to exist around many BCGs is often underestimated, leading to both the total stellar mass and effective radii being similarly affected.  A Hernquist component requires both of these latter parameters and hence would be highly uncertain if anchored from 2MASS data.  Similar uncertainties arise at other wavelengths, hence we removed this model potential from consideration.

Away from its cored region, the SIS is defined entirely by a single velocity dispersion that is constant at all radii.  We therefore utilise the isophotal radii ($r_{\rm K20}$) and apparent magnitudes ($m_{\rm k20}$) reported by 2MASS that are defined as the radius at which the surface brightness reaches 20~mag/sq.''.  This isophotal magnitude should give us a reliable measurement of the light within a certain radius, which can be used to obtain an equivalent stellar velocity dispersion at the same radius.  This $m_{\rm k20}$ was corrected for galactic extinction \cite[][]{Schlegel98}, as well as being evolution and K-corrected following \cite[][]{Poggianti97}. The enclosed stellar mass within $r_{\rm K20}$ is calculated using
\begin{equation}
\log\frac{M}{L_{\rm K}} = -0.206 + 0.135~(B-V)
\end{equation}
\noindent from \citet[][]{Bell03} who derived mass-to-light relations for a large mixed sample of galaxies.  Whilst isolated giant ellipticals are typically `red and dead', BCGs often display enhanced blueness due to ongoing low level star formation \cite[][]{Rafferty08} meaning that a mixed sample is reasonable. Equivalent relations for only BCGs were unavailable.  We adopt a corrected $B-V$ colour of 1.0 that is appropriate for massive BCGs \cite[][]{Baldry08}.  Our derived stellar mass within $r_{\rm K20}$ was converted to a circular velocity and finally an equivalent stellar velocity dispersion using the relation 
\begin{equation} \label{MassEquation}
V_{\rm c} = (1.32 \pm 0.09)\sigma_{*} + (46 \pm 14)
\end{equation}
\noindent from \citet[][]{Pizzella05}, who derived this correlation using H{\sc i} circular velocity measurements for a sample of elliptical galaxies.

This {\em equivalent stellar velocity dispersion} ($\sigma_{*}$) is not a measureable quantity -- it is the inferred velocity dispersion that would be measured at $r_{\rm K20}$ if the BCG consisted only of its stars.  The total halo mass of the cluster is believed to be well accounted for by the NFW profile.  Self similarity of dark-matter haloes suggests that the BCG itself should be considered the dominant galaxy within the cluster halo and is thus not expected to reside within its own dark-matter sub-halo \cite[e.g.][]{George12}.  Combining an NFW component and an isothermal sphere with velocity dispersion equal to our derived $\sigma_{*}$ should result in a mass profile that accounts for the total cluster potential as well as the stellar mass of the BCG at the centre.

To account for the mass of the fixed stellar component when fitting for the cluster NFW potential, we use the \textsc{isonfwmass} model with the isothermal normalisation fixed to match our derived $\sigma_{*}$. The $r_{\rm I}$ parameter (see Equation \ref{IsothermalEquation}) is set to an arbitrarily small but non-zero value ($\ll$1~kpc), ensuring that our cored isothermal component's potential is equivalent to a basic SIS\footnote{Interestingly it was found that the basic SIS model matched the low altitude mass tracers better than a cored-isothermal model with the core radius set to match the stellar core radius.  We speculate that this may suggest that at the smallest radii ($\leq$2~kpc) the mass of the SMBH starts to become important to the total mass, but this is beyond the scope of this work.} model at all radii of interest.  Our input and fitted mass model parameters are listed in Table \ref{ISONFW_fits_table}.  

\subsection{Comparison Mass Tracers}

Our mass profiles are shown in Figures \ref{MassProfiles} (A2029, A2199, A496, A85) and \ref{A780_Mass_Profiles} (Hydra~A).  To ensure that these profiles are reliable we require comparison mass tracers.  All five of our clusters were modelled in \citet[][]{Main15}, who obtained hydrostatic cluster mass estimates from their Chandra observations using the same \textsc{clmass} package of mixing models that we have used here.  With the caveat that our data and methods are therefore not independent, we encouragingly find consistent total cluster masses (reported as M$_{2500}$) with these authors.  Additionally, A2029 was included in the X-ray cluster analyses of both \citet[][]{Vikhlinin06} and \citet[][]{Allen08}.  We plot their M$_{2500}$ values on Figure \ref{MassProfiles} and again find consistency.

X-ray derived cluster masses can be in tension with other tracers, leading us to search for otherwise obtained mass estimates.  \citet[][]{Okabe12} used weak-lensing analysis to derive an outer mass profile for Hydra~A, which we show in Figure \ref{A780_Mass_Profiles} to be in agreement with our X-ray measurements.  This profile was in similar agreement with the X-ray mass profile of \citet[][]{David01}. A2029 and A85 were included in the weak-lensing analysis of \citet[][]{Cypriano04}.  These authors approximated the cluster mass on large ($\gtrsim$arcmin) scales as an isothermal sphere and reported cluster velocity dispersions.  Although a singular isothermal sphere (SIS) is known to be a poor fit to cluster mass profiles across all radii \cite[][]{Schmidt07}, this is a reasonable approximation at the scales of interest.  We converted their velocity dispersions at R$_{2500}$ to masses using the standard SIS potential: $\Phi_{\rm iso}(r)$~=~2~$\sigma^{2}~\ln$(r).  \citet[][]{Kubo09} used SDSS data \cite[][]{Abazajian09} to study weak-lensing signatures around seven nearby clusters, including A2199, for which they report an M$_{200}$.  Our Chandra coverage does not extend to these altitudes and hence our profiles are highly uncertain if extrapolated so far (see Section \ref{SystematicSection}).  Instead, we use the assumption of \citet[][]{Cypriano04} that the cluster outskirts can be approximated as a SIS to convert this M$_{200}$ to an M$_{2500}$.  As an additional test of the uncertainty of this extrapolation we adopt the 363.0~kpc NFW scale radius of \citet[][]{Main15} for this cluster and vary the normalisation to match the M$_{200}$ reported in \citet[][]{Kubo09} with an NFW profile, from which we then read an equivalent M$_{2500}$.  These two M$_{2500}$ values are consistent, in addition to being in agreement with our derived mass profile.  Although a mass estimate from weak-lensing could not be found for A496, this cluster does appear in the sample of clusters to which \citet[][]{Wojtak10} fitted mass profiles using the dynamics of the constituent galaxies.  All of these weak-lensing and dynamical total cluster mass estimates are shown on Figure \ref{MassProfiles} and found to be in agreement with our mass profiles.

The five clusters within this paper were selected as having tracers of their inner mass profiles at various radii.  For Hydra~A these come from the IFU observations of cold gas motions in the central galaxy presented by \citet[][]{Hamer14}.  We also find three stellar velocity dispersions \cite[][]{Heckman85} reported in the HyperLEDA database \cite[][]{Makarov14} for the central galaxy in the Hydra~A cluster, which we convert to mass estimates and plot at a representative 10~kpc.  The remaining four clusters all have stellar velocity dispersion profiles featured in \citet[][]{Fisher95}, which we convert to estimates of gravitating mass using Equation \ref{MassEquation}, and plot on Figure \ref{MassProfiles}.  One caveat is that converting stellar velocity dispersion profiles to inferred masses assumes an isotropic distribution with minimal net flow.  Ordered, and disordered but anisotropic, stellar motions within the BCGs can therefore bias the resultant inferred masses.  Indeed, rotating H$\alpha$ gas clouds are seen in some BCGs \cite[][]{Hamer16} so it is perhaps not unreasonable to expect some degree of ordered motion also in the stellar component.  However we note that \citet[][]{Fisher95} measured very little rotation in their sample of 13 BCGs.  It is also known from IFU surveys that amongst ellipticals there is an anti-correlation between mass and rotational support, with all `non-rotators' in the SAURON survey having M~$>$~10$^{11.25}$~M$_{\odot}$ \cite[][]{Emsellem07}.  Furthermore, whilst not all slow-rotators are massive, all massive galaxies tend to be slow-rotators \cite[][]{Emsellem11}.  Effectively this means BCGs are almost always `non' or `slow rotators'.  Additionally, \citet[][]{Loubser08} measured the anisotropy parameter \cite[][]{Kormendy82} as a function of luminosity and showed BCGs are even less rotationally supported than field giant ellipticals, which as a class already show very low levels of rotation. The evidence therefore suggests that stellar velocity dispersions are a reasonable mass proxy for BCGs at the centres of galaxy clusters.

Overall we find good agreement between our calculated mass profiles and comparison tracers at both large and small radii, thus showing this technique to be a feasible approach for tracing cluster mass profiles down into the central galaxy.

\subsection{Estimating Uncertainties}
\subsubsection{Random Statistical Errors}
Error estimation within \textsc{xspec} itself defaults to Gaussian statistics.  The parameters describing the NFW scale radius ({\em nfwa}) and potential ({\em nfwpot}) in the mass mixing models \cite[see][]{Nulsen10} are co-dependent and hence this approach will give unreliable uncertainties.  Instead, to estimate errors on these we create a 5000 iteration \textsc{mcmc} chain, which is used to obtain the reported 1-$\sigma$ errors (see Table \ref{ISONFW_fits_table}).  Uncertainty on the 2MASS magnitudes are propagated through to $\sigma_{*}$ and a similar 5000 iteration \textsc{mcmc} chain is then created for the isothermal component.  Uncertainties on M$_{2500}$, and at each step of the mass profiles, are estimated by running 5000 combinations of these chains.

\subsubsection{Systematic Error Consideration} \label{SystematicSection}
A potential systematic effect in our adopted approach concerns the isothermal component continuing to increase to large radii ($\geq$20~kpc), beyond where we would reasonably expect the stars of the BCG to form a significant contribution to the total mass budget.  This problem could be alleviated by truncating the isothermal component's mass with a sharp cut at some radius, although this approach is rejected as it would introduce an unphysical discontinuity in the density and associated mass profiles.  Alternatively a maximum mass could be assigned to the isothermal component, and a damping term included so that the profile asymptotes to this maximum mass rather than having a sharp discontinuity.  However, this approach suffers from the issue that damping the isothermal component would cause the resultant mass at $r_{\rm K20}$ to be less than our calculated stellar mass at that radius. Furthermore, a physically motivated maximum mass and damping radius would have to be associated with the highly uncertain BCG effective radii R$_{\rm e}$.  

\begin{figure}
	\includegraphics[width=\columnwidth]{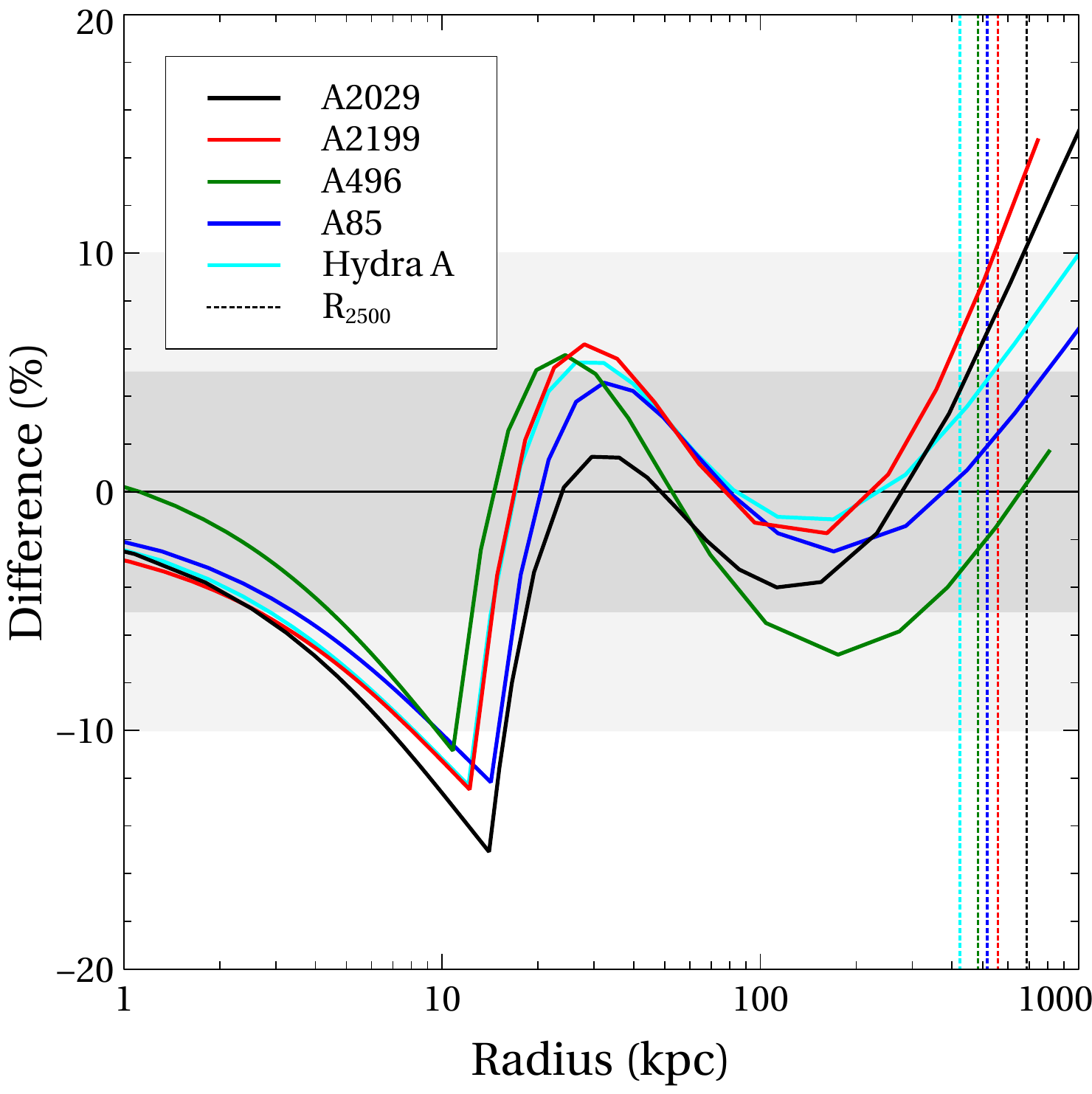}
    \caption{Percentage difference (see Equation \ref{DifferenceEquation}) in mass between our \textsc{isonfwmass} profiles, and a singular NFW fit where a constant mass equivalent to the isothermal component at 10~kpc is accounted for in fitting.  The vertical dotted lines correspond to the R$_{2500}$ for each colour-coded cluster.  We see systematic differences in the resultant mass profiles are $\lesssim$10\% out to at least R$_{2500}$, beyond which the fits diverge. The sharp transition at r$\approx$10~kpc is due to the artifical truncation of the isothermal component in the singular NFW fit.  For a full description see section \ref{SystematicSection}.}
    \label{SystematicEffect}
\end{figure}

The mass distribution associated with the NFW potential (Equation \ref{Equation:NFWPotential}) is
\begin{equation}
M(r) = 4\pi\rho_{0}r_{\rm s}^{3} \Bigg[ \ln\bigg(1 + \frac{r}{r_{\rm s}}\bigg) - \frac{r/r_{\rm s}}{1+r/r_{\rm s}} \Bigg]    
\end{equation}
where $r_{\rm s}$ is the scale radius.  At large $r$ ($r{\gg}r_{\rm s}$) the NFW component's mass therefore varies proportionally to $\ln(r/r_{\rm s})-1$.  For the singular isothermal sphere potential $2\sigma^2\ln(r/r_{\rm I})$, the mass distribution $=2\sigma^{2}r/G$.  Hence at sufficiently large radii, the non-truncated isothermal componenent will always dominate the logarithmic NFW mass distribution.  However, the radial range that we are interested in is typically less than a few NFW scale radii and so we are less affected by this effect. At radii below the NFW scale radius this component's mass distributions increases more quickly than that of the isothermal at the same radii.

The problem of the isothermal component existing in our fits beyond where the stars dominate is mitigated because the mass associated with the NFW component rises much faster over the radial range in which we are interested than the mass associated with the isothermal component.  Indeed, this latter component contributes an increasingly smaller fraction of the mass budget towards the largest radii that we trace our mass profiles to. Nevertheless, although the \textsc{isonfwmass} model with fixed isothermal component finds the optimised NFW component to provide the best-fit total mass profile, having an artifically inflated contribution from the isothermal component that becomes fractionally less important to larger radii could affect the shape of our final mass profiles.  Practically this means less mass is `free' at smaller radii to be fitted with the NFW.  Accordingly, the NFW scale radius R$_{\rm s}$ is pushed to larger values so that the NFW mass contribution is lessened towards the centre, whilst the normalisation is increased to recover the total mass at higher radii.  This effect is likely to subtly change the shape of the final mass profile, potentially most prevalently at the intermediate radii (few 10s~$\lesssim$~r~$\lesssim$~few 100s kpc) where lower entropy core gas is postulated to be lifted and become unstable in the stimulated feedback model.

We can quantify this effect by looking at the fractional difference with radius between the final mass profiles from our \textsc{isonfwmass} fits, and a comparison mass profile consisting of a fitted NFW profile coupled with an isothermal component truncated at some radius.  To obtain these comparison profiles we first measure the isothermal component's mass at 10~kpc for each cluster.  We then insert this central mass as a fixed constant, and fit an NFW profile to account for the remaining (majority) mass in each cluster.  Note that the central region ($\gtrsim$10kpc) is excluded in each of these comparison fits.  The comparison profiles therefore consist of the isothermal component truncated at 10~kpc, coupled with the NFW component that was fitted outside of this radius but accounting for this constant central mass.  The resulting profiles are therefore effectively produced similarly to our \textsc{isonfwmass} fits but with an artificially truncated isothermal component -- an approach that was earlier rejected since it introduces an unphysical mass discontinuity at the truncation radius.  Nevertheless, this approach is suitable for our current investigation of the systematic effect of a non-truncated isothermal component.  

\begin{figure*}    
  \centering
    \subfigure{\includegraphics[width=8cm]{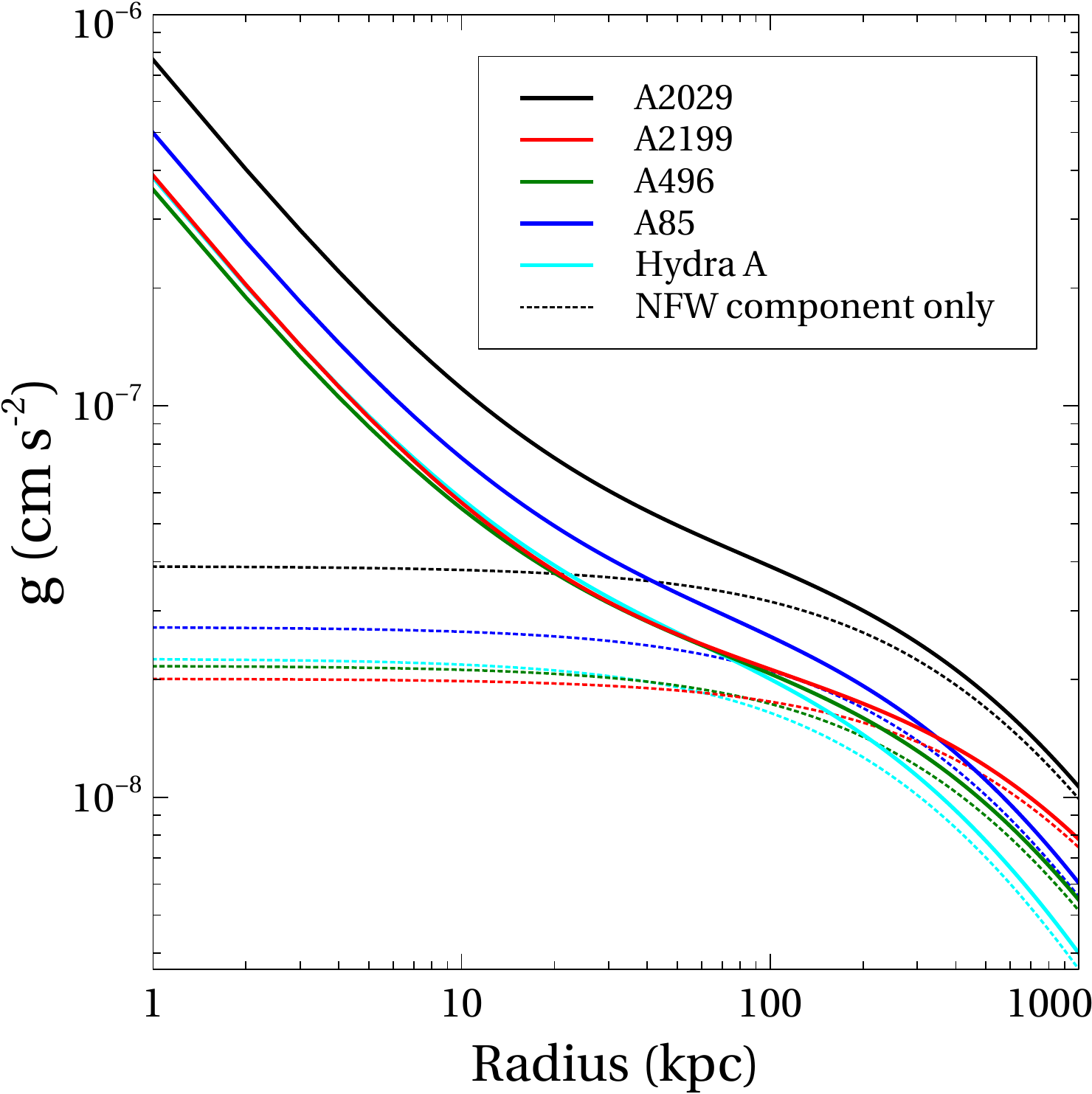}}
    \subfigure{\includegraphics[width=8cm]{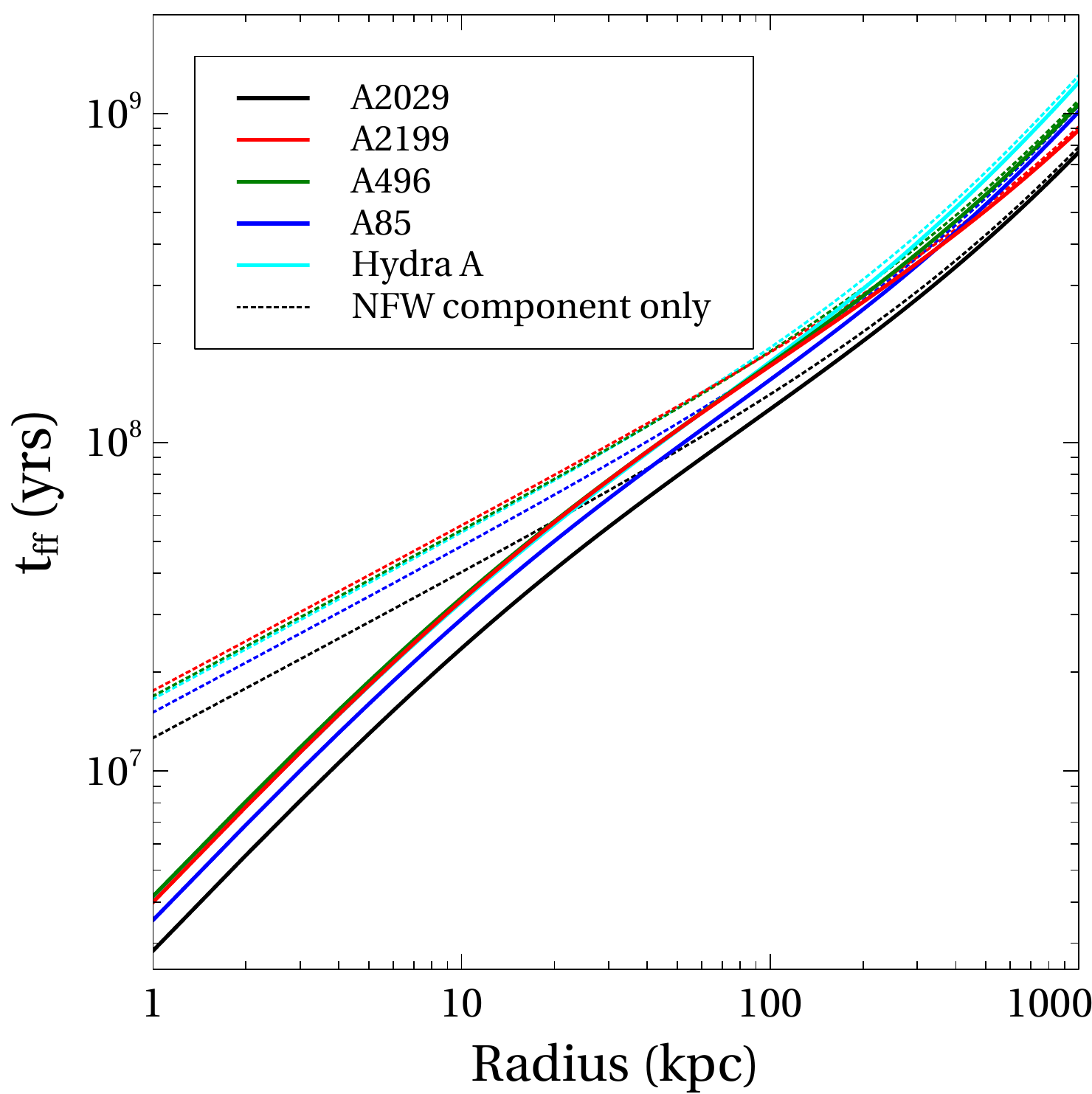}}
  \caption{Radial run of acceleration (left) and free-fall time (right) corresponding to the mass profiles derived in Section \ref{Mass_Profiles}.  Shown as dotted lines are the equivalent parameters when the additional isothermal mass component associated with the BCG is neglected.  The acceleration felt by cold gas clumps in cluster centres, as seen by ALMA, is highly dependent upon their distance from the central galaxy, with implications for the mechanisms required to retard the velocities of these clouds within the stimulated feedback model.  Note also that the discrepancy of 2-5 in free-fall time when the isothermal component is neglected is similarly dependent upon radial distance from the assigned centre, highlighting the importance of the choice of cluster centre in the ultimate measurement of the (\tctff)$_{\rm min}$.}
 \label{FreeFallProfile}
\end{figure*}

The percentage difference between these profiles and our final \textsc{isonfwmass} profiles, defined as
\begin{equation} \label{DifferenceEquation}
{\rm Difference} = 100 \times \Bigg(1 - \frac{M_{\rm comparison}}{M_{\rm isonfwmass}}\Bigg)
\end{equation}
is shown in Figure \ref{SystematicEffect}.  Each of the ratios in this plot tells the same tale.  The \textsc{isonfwmass} and comparison profiles converge towards $r=0$ as the NFW component becomes increasingly negligible.  Below $\sim$10~kpc the \textsc{isonfwmass} fit gives less mass, which is consistent with it tending to have a higher NFW scale radius and thus a smaller NFW contribution at smaller radii.  A sharp transition is seen at $\approx$10~kpc where the isothermal component in the comparison profiles is artificially truncated.  We reiterate that the isothermal component's truncated mass is still accounted for in the comparison profile at all radii, just as a fixed mass beyond the truncation radius.  The sharp transition is because the isothermal component still constitutes a large fraction of the total mass at $\sim$10~kpc.  The \textsc{isonfwmass} model therefore continues to rise at almost the same rate beyond the truncation radius, whereas the comparison profile barely increases at radii just above r$\sim$10~kpc, causing a sharp (but not instant) transition from negative to positive values in Figure \ref{SystematicEffect}.  The NFW component rises more quickly than the isothermal component at radii much less than the NFW scale radius.  However, at $\sim$20--40~kpc the \textsc{isonfwmass} profile gives slightly higher mass because its still rising isothermal component continues to contribute an appreciable fraction of the mass budget.  At radii $\gtrsim$40~kpc the NFW component begins to dominate the mass budget.  Between $\sim$40--300~kpc the larger NFW scale radius in the \textsc{isonfwmass} fits means that this model's mass prediction is systematically lower than that of the comparison.  Finally, the lower scale radius of the comparison model causes its mass profile to flatten at $\sim$300~kpc whereas the \textsc{isonfwmass} continues to rise, causing the profiles to diverge.  The different mass scalings of the NFW and isothermal components at large r, as discussed near the beginning of this section, will cause this divergence to grow to yet larger radii.

There are clearly systematic uncertainties associated with the inclusion of a non-truncated isothermal component.  However, these result in only around $\lesssim$10\% difference in the mass profiles out to at least R$_{2500}$.  The free-fall time t$_{\rm ff}\propto$M$^{-1/2}$ hence this is effect is further lessened.  Our derived profiles are therefore believed to be robust to at least this accuracy below R$_{2500}$, although we caution that extrapolation of our fits to greater radii than this is likely to be uncertain.

\subsection{The Importance of Defining the Cluster Centre} \label{CentreSection}

A major motivation of our work is to understand the role of mass in maintaining the balance between heating and cooling in cluster cores.  As previously mentioned, recent simulations have suggested that one important facet of this may be a cooling instability that triggers hot gas to condense at some threshold value of \tctff~ \cite[e.g.][]{McCourt12,Sharma12b,Gaspari13,Prasad15,Li15}.  Observationally testing such a threshold is challenging.  In Section \ref{Section:ResolutionEffects} we investigated the effect of resolution on measured $t_{\rm cool}$.  Being able to trace the mass, and associated parameters such as acceleration and free fall time, to small radii is similarly crucial in being able to test these models.

A common approximation for the free-fall time \cite[e.g.][]{Mcdonald15,Voit15a}, defined with respect to to local gravitational acceleration g, is 
\begin{equation}
t_{\rm ff} = \sqrt{\frac{2r}{g}} .  \label{equation:tff}
\end{equation}
We adopt this definition of the free-fall time for our profiles in order that they are easily comparable to the literature.

Figure \ref{FreeFallProfile} shows the radial run of accelerations and free-fall times corresponding to the mass profiles derived in the previous section.  Also shown are the equivalent parameters if the isothermal component of the mass is neglected (i.e. for the cluster NFW only).  The accelerations associated with these NFW-only mass profiles are in good agreement with the NFW-only acceleration profile derived for PKS0745-191 in \citet[][]{Sanders14}.  However, consideration of Figures \ref{MassProfiles} \& \ref{A780_Mass_Profiles} shows the requirement to account for the BCG, with Figure \ref{FreeFallProfile} highlighting that the calculated gravitational acceleration can otherwise be underestimated by more than an order of magnitude.

The importance of the assigned cluster centre is apparent when considering the right-hand panel of Figure \ref{FreeFallProfile}, even amongst this small exploratory sample.  There is a factor of 2-5 difference in inferred free-fall time between the NFW-only component and the total mass component.  As the free-fall time is dependent upon distance from the centre, the choice of dynamical centre can have a large effect on the ultimate minimum value of, for example the \tctff~ that is measured.  

Defining the centre of a cluster is an imperfect art.  Even the assertion that there is a centre to be defined implies a degree of symmetry that is not always present. However, the importance of this latter point is mitigated here since the most disturbed clusters are by and large those with long central cooling times.  A factor of a few difference in \tctff~ is therefore less important since for these clusters cooling is not seen and the measured \tctff~ is usually at least an order of magnitude above any postulated cooling thresholds \cite[e.g.][]{Voit15a}.  Conversely, the clusters where cooling is expected to occur have not recently experienced major disturbance and the assumption of spherical symmetry is typically reasonable, especially on large scales.  Nonetheless, as seen from Figure \ref{FreeFallProfile} a mis-centering by only $\sim$10~kpc could greatly affect the \tctff.

Multiple definitions exist for the cluster centre.  Perhaps the most natural is the peak of the cluster potential, as is often defined in simulations.  Similarly, lensing mass measurements use reconstructive methods to infer the mass distribution, with the centre typically taken as the potential peak.  However, often this has large errors due to intrinsic galaxy shapes and cluster sub-structure where even small centroid offsets can cause cluster mass to be underestimated by $\sim$30\% \cite[][]{George12}.  

\begin{figure*}    
  \centering
    \subfigure{\includegraphics[width=17cm]{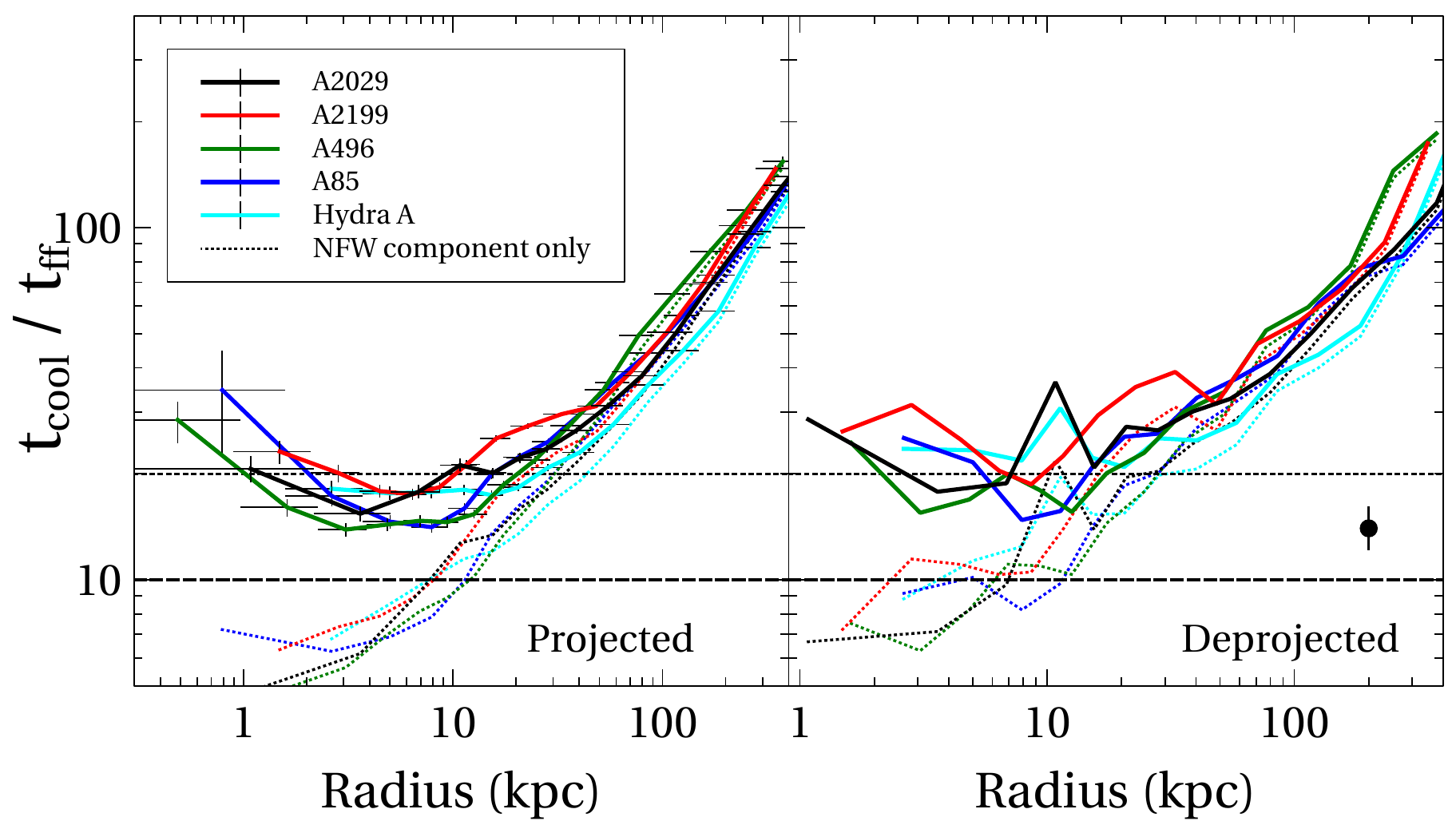}}
  \caption{Projected (left) and deprojected \tctff~ profiles for our evaluation sample.  Dotted lines on both panels shows the equivalent parameters if the isothermal mass component associated with the BCG is neglected.  Horizontal error bars on the left-hand panel reflect bin width.  The same bin widths are used on the right-hand panel but not shown.  Similarly, only a representative error bar is shown in the right-hand panel to aid clarity.  Minimum values of \tctff~ lie in the range 14.8--21.8 for our deprojected profiles.}
 \label{tctffProfiles}
\end{figure*}

With X-ray data the two most natural choices of centre are the X-ray centroid and the X-ray peak.  Usually, but not always, these indicators are in close agreement \cite[e.g.][]{Mann12}.  Another natural choice for the cluster centroid is the middle of the BCG.  Although not all clusters contain a BCG, the vast majority of cool core clusters do. Furthermore, these BCGs are typically located close ($\lesssim$50~kpc) to the X-ray peak and/or centroid \cite[e.g.][]{Lin04,Mann12,Rozo14,vonderLinden14}.  Indeed, studying a sample of 129 massive groups \citet[][]{George12} found the massive galaxy closest ($\leq$75~kpc) to the X-ray centroid (i.e. the BCG) to be the best tracer of dynamical centre out of 8 tracers considered, a result supported by the simulations of \citet[][]{Cui15}. In a sample of 19 clusters, \citet[][]{Loubser16} found that the four with BCGs located within a projected 5~kpc of the X-ray centroid were the only BCGs in their sample with appreciable ongoing star formation.

Returning to the left-hand panel of Figure \ref{FreeFallProfile}, the acceleration associated with the NFW component of our mass distribution is seen to be relatively flat inwards of about 100~kpc.  This is due to the r$^{-2}$ dependency of the gravitational acceleration being counteracted by the NFW density slope approaching an $r^{-1}$ scaling well within the scale radius such that M$\propto$r$^{2}$.  The effect of the galaxy itself (seen via the isothermal component) is most potent at radii $<$10~kpc.  The cooling instability models are concerned primarily with gas condensing from the hot phase and ultimately fuelling AGN feedback \cite[e.g.][]{McCourt12,Sharma12b,Prasad15}. That the BCGs in cool core clusters are found to be typically $<$50~kpc from the cluster centre, and that the local gravity associated with the cluster halo on these scales is less dependent on position than gravity associated with the BCG itself therefore compels us to take the centre of the BCG as the most appropriate centroid for our study.  This choice is further supported by the distribution of molecular gas seen with ALMA, which is found to lie near the BCG and be associated with X-ray cavities that are directly associated with it \cite[e.g.][]{McNamara14,Russell14,Russell16,Tremblay16}.

We note retrospectively that such a criterion was used for each of the clusters in this current study, where in each case a very clear BCG exists in close association with the X-ray centroid (see also Section \ref{Section:SpectralExtraction}).  However, in non cool core clusters the identification of a BCG is often uncertain, and that the more diffuse X-ray atmosphere has a less well-defined centroid.  In all cases for the larger sample we therefore suggest that the cluster centre be assigned a ranked centring grade.  For clusters with a clearly dominant optical/IR BCG that resides close to the X-ray centroid then the centre of this galaxy be assigned as cluster centre.  In clusters with no clear BCG then the X-ray centroid itself is taken to be the cluster centre, with the closest major galaxy classified as the BCG.  The inclusion of an isothermal component to account for this ``BCG'' would overestimate the mass in the inner regions of these clusters -- thus where no central BCG is present a velocity dispersion of zero is appropriate, and the fits revert to a single NFW.

\section{Discussion} \label{Section:FinalRemarks}

One of our main motivations is to understand AGN feedback as a function of mass \cite[][]{Main15}.  Accurate mass profiles that extend down into cluster centres is important because the processes governing gas cooling within cluster cores are believed to be underpinned by dependence on the cluster mass.  For example, it has long been known \cite[e.g.][]{Cowie80,Nulsen86} that gas becomes thermally unstable when \tctff~ falls to unity.  More recently, precipitation models suggest that gas condensation from the hot phase may occur below a threshold value of \tctff$\sim$10 \cite[e.g.][]{McCourt12,Sharma12b,Gaspari13,Prasad15}.  Previous studies \cite[e.g][]{Gaspari12,Voit15a} found \tctff~ profiles reach a minimum value at $\sim$5-10~kpc before increasing to yet smaller radii.  Again this highlights the importance of accurately determining the central mass profiles and having deprojected density and temperature measurements down to small radii.

A full discussion of gas cooling in cluster cores in the context of the precipitation and stimulated feedback models is beyond the scope of this paper and therefore left to upcoming work with a much larger cluster sample.  However, as a proof of concept Figure \ref{tctffProfiles} shows the projected and deprojected \tctff~ ratios for our exploratory sample.  Note that a 10\% systematic mass uncertainty (Section 4.5.2) is factored into our free-fall time error calculations. Amongst our small sample we find minimum values of \tctff~ that span the range 14.8--21.8 (Figure \ref{tctffProfiles}).  Whilst this sample is far too small to draw any strong conclusions about the spread of \tctff~ minima in the general population, we note that none of these fall below the postulated nominal instability threshold of 10, and are therefore possibly in tension with precipitation models.  It is worth noting that the unusual cluster A2029, which falls below both the cooling time and entropy thresholds, but fails to exhibit H$\alpha$, does not stand out in this sample, with (\tctff)$_{\rm min}$ = 17.8.  In the stimulated feedback model \cite[][]{McNamara16} this is attributed to A2029's inability to lift low entropy gas from the core to an altitude at which gas stability is breached.

Also shown in Figure 9 are the \tctff~ profiles found when the BCG (i.e. the isothermal component) is neglected.  It is evident that both the minimum value of \tctff~ and the radius where it is found are dictated by the BCG. Deriving an accurate central mass profile (and subsequently \tctff) therefore requires a measurement of the central galaxy's mass that is tailored to each individual cluster, as is done in the technique described within this paper.

A final point that can be illustrated using Figure \ref{tctffProfiles}, and also Figure \ref{DeprojectedProfiles}, is the importance of deriving fully deprojected cluster properties such as density, and in particular temperature, down to small radii.  As discussed in Section \ref{DeprojectionSection}, our deprojected entropy profiles do not show any suggestion of flattening towards an ``entropy floor''.  Instead our central entropies follow a power-law shape and continue to drop below $\sim$10~keV.  These central entropies are all lower than the equivalent values for these clusters in the ACCEPT database \cite[][]{Cavagnolo09} but are consistent with the fully deprojected findings of both \citet[][]{Panagoulia14} and \citet[][]{Lakhchaura16}.  These latter authors highlighted that the ACCEPT profiles combined projected temperatures with higher resolution deprojected densities, and concluded that the higher temperatures therefore cause the higher entropy.  As can be seen by comparison of Figures \ref{ProjectedProfiles} and \ref{DeprojectedProfiles}, the same effect causes our own lower entropies since we find lower central temperatures after deprojection.  Projection effects are increasingly important towards the centre and therefore projected temperature is also likely to be increasingly over-estimated towards the centre (see also Section \ref{Section:ResolutionEffects}), perhaps explaining the apparent flattening of entropy profiles as reported previously.

In Figure \ref{tctffProfiles} the minimum values of \tctff~ are clearly well sampled radially.  However, as discussed in Section \ref{Section:ResolutionEffects}, limited angular resolution can often lower the accuracy to which central cluster properties can be determined.  As a test, we can use our coarsely binned spectra from Section \ref{Section:ResolutionEffects} to simulate the effect that lower spatial resolution would have on the measured minimum values of \tctff~ displayed in Figure \ref{tctffProfiles}.  Such lower spatial resolution could arise due to clusters being located at higher redshift, or alternatively be due to shallower data requiring that larger central annuli be used to obtain enough X-ray counts that deprojected models can successfully be fitted.  We calculated \tctff~ profiles for our coarsely binned spectra (see Section \ref{Section:ResolutionEffects}) and in all cases found that the measured deprojected (projected) \tctff~ minima were a factor of 2.6--9.4 (2.0--6.4) higher than when finer sampling was used.  In conjunction with our findings in Section \ref{Section:ResolutionEffects} that extrapolation of cooling profiles becomes uncertain below the radii at which temperature can be directly measured, the most accurate \tctff~ investigations can therefore only be performed on clusters whose central $\sim$5--10~kpc are well sampled, corresponding to z$\lesssim$0.6 with Chandra.  Count rates will of course practically restrict this range further.

\section{Conclusions} \label{Section:Conclusions}

In this paper we have used a small exploratory sample of clusters with deep Chandra observations and ancillary mass measurements to develop a method for determining mass distribution across a wide radial range.  Additionally, we have considered a number of potential observational effects that could impact measurements of cooling properties in galaxy clusters.  In particular we find that:

\begin{itemize}
    \item Mass distributions can be well modelled by a combination of a singular isothermal sphere component anchored to the stellar mass of the BCG, and an NFW profile fitted to the X-ray data.  The effect of the isothermal component continuing to large radii gives $\lesssim$10\% uncertainty on the final mass profile out to M$_{2500}$.  
    \item Entropy profiles are best described by a power-law down to $\sim$1~kpc, in agreement with \cite[][]{Panagoulia14}
    \item We argue that, for clusters containing an obvious central BCG, the BCG represents the best location for the cluster centre.
    \item Tentative evidence suggests that the cooler temperature component in a two-temperature thermal model fit to central ICM may be associated with extended nebular emission.  A single-temperature fit is found to be adequate to trace cooling of the hot phase down to $\sim$1~kpc.
    \item The central regions of galaxy clusters (r~$\lesssim$~20~kpc) are poorly modelled by a constant temperature.  Fixing temperature within the inner $\sim$20~kpc leads to errors of $\sim$10\% on cooling time, which increase to small radii.
    \item None of our exploratory sample have \tctff~ below 10, although 4/5 exhibit H$\alpha$ emission indicative of ongoing cooling.
\end{itemize}

The techniques described within this paper are to be applied to a large sample of clusters to further investigate thermal instability and AGN feedback within galaxy clusters from an observational viewpoint in upcoming papers (Hogan {\em et al. in prep.}, Pulido {\em et al. in prep.}).

\acknowledgements

Support for this work was provided in part by the National Aeronautics and Space Administration through Chandra Award Number G05-16134X issued by the Chandra X-ray Observatory Center.   MTH, BRM, ANV, and FP acknowledge support from the Natural Sciences and Engineering Research Council of Canada.  HRR acknowledges support from ERC Advanced Grant Feedback 340442.  The scientific results reported in this article are based on observations made by the Chandra X-ray Observatory and has made use of software provided by the Chandra X-ray Center (CXC) in the application packages CIAO, ChIPS, and Sherpa.  The plots in this paper were created using Veusz.

\bibliographystyle{apj}
\bibliography{refs}

\end{document}